\documentclass[acmsmall]{acmart}
\usepackage{cleveref}
\usepackage{tikz}
\usepackage{listings}
\usepackage{color}
\usetikzlibrary{arrows, arrows.meta, automata, calc, shapes, positioning, fit}

\definecolor{darkblue}{rgb}{0,0,0.5}
\definecolor{darkgreen}{rgb}{0,0.3,0}
\definecolor{darkpink}{rgb}{0.4,0,0.3}
\definecolor{graygreen}{rgb}{0.3,0.5,0.3}
\definecolor{grayblue}{rgb}{0.2,0.2,0.6}
\definecolor{grayred}{rgb}{0.5,0.2,0.2}

\AtBeginDocument{%
  }

\acmJournal{PACMPL}
\acmVolume{1}
\acmNumber{ICFP} % CONF = POPL or ICFP or OOPSLA
\acmArticle{1}
\acmYear{2024}
\acmMonth{9}
\acmDOI{} % \acmDOI{10.1145/nnnnnnn.nnnnnnn}
\startPage{1}

\setcopyright{none}
\begin{document}

\newcommand{\ambient}{\textsc{Ambient-Verifier}}
\newcommand{\aslTranslator}{\textsc{ASL-Translator}}
\newcommand{\cerridwen}{\textsc{Cerridwen}}
\newcommand{\crucible}{\textsc{Crucible}}
\newcommand{\dismantle}{\textsc{Dismantle}}
\newcommand{\elfEdit}{\textsc{ELF-Edit}}
\newcommand{\flexdis}{\textsc{Flexdis86}}
\newcommand{\grease}{\textsc{GREASE}}
\newcommand{\grift}{\textsc{GRIFT}}
\newcommand{\macaw}{\textsc{Macaw}}
\newcommand{\macawAArch}{\textsc{Macaw-AArch32}}
\newcommand{\macawAArchSymbolic}{\textsc{Macaw-AArch32-Symbolic}}
\newcommand{\macawBase}{\textsc{Macaw-Base}}
\newcommand{\macawPPC}{\textsc{Macaw-PPC}}
\newcommand{\macawPPCSymbolic}{\textsc{Macaw-PPC-Symbolic}}
\newcommand{\macawRISCV}{\textsc{Macaw-RISCV}}
\newcommand{\macawRefinement}{\textsc{Macaw-Refinement}}
\newcommand{\macawSemMC}{\textsc{Macaw-SemMC}}
\newcommand{\macawSymbolic}{\textsc{Macaw-Symbolic}}
\newcommand{\macawX}{\textsc{Macaw-x86}}
\newcommand{\macawXSymbolic}{\textsc{Macaw-x86-Symbolic}}
\newcommand{\mcTrace}{\textsc{MCTrace}}
\newcommand{\pate}{\textsc{PATE}}
\newcommand{\reopt}{\textsc{Reopt}}
\newcommand{\refurbish}{\textsc{Refurbish}}
\newcommand{\renovate}{\textsc{Renovate}}
\newcommand{\saw}{\textsc{SAW}}
\newcommand{\semmc}{\textsc{SemMC}}
\newcommand{\semmcAArch}{\textsc{SemMC-AArch32}}
\newcommand{\semmcPPC}{\textsc{SemMC-PPC}}
\newcommand{\surveyor}{\textsc{Surveyor}}
\newcommand{\TIE}{\textsc{TIE}}
\newcommand{\whatFour}{\textsc{What4}}
\newcommand{\xEightySix}{x86-64}

\newcommand{\todo}[1]{}
\renewcommand{\todo}[1]{{\color{red} TODO: {#1}}}

\tikzstyle{nodeStyle}=[
    draw = black,
    thick,
    minimum size = 4mm,
    rectangle,
    rounded corners,
    inner sep = 2mm,
    align=center
]

\tikzstyle{edgeStyle}=[
    ->,
    align=center
]

\title{\macaw: A Machine Code Toolbox for the Busy Binary Analyst}

\author{Ryan G. Scott}
\affiliation{%
  \institution{Galois, Inc.}
  \country{United States}}
\email{rscott@galois.com}

\author{Brett Boston}
\affiliation{%
  \country{United States}}
\email{brett@brett.boston}

\author{Benjamin Davis}
\affiliation{%
  \institution{Galois, Inc.}
  \country{United States}}
\email{ben@galois.com}

\author{Iavor Diatchki}
\affiliation{%
  \institution{Galois, Inc.}
  \country{United States}}
\email{diatchki@galois.com}

\author{Mike Dodds}
\affiliation{%
  \institution{Galois, Inc.}
  \country{United States}}
\email{miked@galois.com}

\author{Joe Hendrix}
\affiliation{%
  \country{United States}}
\email{joe.d.hendrix@gmail.com}

\author{Daniel Matichuk}
\affiliation{%
  \institution{Galois, Inc.}
  \country{United States}}
\email{dmatichuk@galois.com}

\author{Kevin Quick}
\affiliation{%
  \institution{Galois, Inc.}
  \country{United States}}
\email{kquick@galois.com}

\author{Tristan Ravitch}
\affiliation{%
  \country{United States}}
\email{tristan@ravit.ch}

\author{Valentin Robert}
\affiliation{%
  \institution{Galois, Inc.}
  \country{United States}}
\email{val@galois.com}

\author{Benjamin Selfridge}
\affiliation{%
  \institution{Galois, Inc.}
  \country{United States}}
\email{benselfridge@galois.com}

\author{Andrei Ștefănescu}
\affiliation{%
  \country{United States}}
\email{andrei@stefanescu.io}

\author{Daniel Wagner}
\affiliation{%
  \country{United States}}
\email{me@dmwit.com}

\author{Simon Winwood}
\affiliation{%
  \country{United States}}
\email{simonjwinwood@gmail.com}

\renewcommand{\shortauthors}{Scott et al.}

\begin{abstract}
When attempting to understand the behavior of an executable, a binary analyst
can make use of many different techniques.
These include program slicing, dynamic instrumentation, binary-level rewriting,
symbolic execution, and formal verification, all of which can uncover insights
into how a piece of machine code behaves.
As a result, there is no one-size-fits-all binary analysis tool, so a binary
analysis researcher will often combine several different tools.
Sometimes, a researcher will even need to design new tools to study problems
that existing frameworks are not well equipped to handle.
Designing such tools from complete scratch is rarely time- or cost-effective,
however, given the scale and complexity of modern instruction set
architectures.

We present \macaw{}, a modular framework that makes it possible to rapidly
build reliable binary analysis tools across a range of use cases.
Statically typed functional programming techniques are used pervasively
throughout \macaw---these range from using functional optimization passes to
encoding tricky architectural invariants at the type level to statically check
correctness properties.
The level of assurance that functional programming ideas afford us allow us to
iterate rapidly on \macaw's development while still having confidence that the
underlying semantics are correct.

Over a decade of development, we have used \macaw{} to support an industrial
research team in building tools for machine code--related tasks.
As such, the name ``\macaw'' refers not just to the framework itself, but also
a suite of tools that are built on top of the framework.
We describe \macaw{} in depth and describe the different static and dynamic
analyses that it performs, many of which are powered by an SMT-based symbolic
execution engine.
We put a particular focus on interoperability between machine code and
higher-level languages, including binary lifting from x86 to LLVM, as well
verifying the correctness of mixed C and assembly code.

\end{abstract}

\maketitle

\section{Introduction}\label{sec:introduction}

\emph{Binary analysis} refers to not one class of tool, but many.
Starting with an unknown binary, there are many tasks that an engineer may want
to accomplish.
These may include \emph{discovering} the structure of the binary,
\emph{transforming} the binary to achieve some goal, \emph{analyzing} the binary
to identify particular properties or vulnerabilities, and/or \emph{formally
verifying} that the binary matches some specification.
These broad tasks can be combined and specialized into many domains, objectives,
tool types, and instruction sets.
The result is that there is as yet no optimal binary analysis tool, but rather a
large space of potentially useful designs.

We present \macaw{}, a binary analysis toolkit which we have built to help us
explore this design space of tools.
\macaw{} exists in a similar niche as frameworks such as Angr~\citep{angr} and,
to a lesser extent, reverse-engineering suites such as Ghidra~\citep{ghidra} and
Binary Ninja~\citep{binja}.
What sets \macaw{} apart are the intended use cases.
\macaw{} is not a standalone tool, but it is instead a Haskell library that can
be used to rapidly construct and evaluate binary analysis tools.
The main \macaw{} library is also accompanied by a suite of libraries for
disassembly, representing architecture semantics, and symbolically executing
machine code, which we collectively refer to as the \emph{\macaw{} ecosystem}.
\macaw{} is designed to help us build quickly, maximize reuse of existing
components, and avoid costly errors during prototyping.
This imposes a particular set of design constraints on the \macaw{} ecosystem,
which we explain in this paper.

The \macaw{} ecosystem was designed from the onset with statically typed
functional programming techniques in mind.
This can be seen in the design of \macaw's intermediate representation (IR),
which encodes information about the semantics of assembly operations and other
architecture considerations at the type level.
\macaw's IR needs to be rigid enough to rule out obviously bad machine-code
programs while flexible enough to bootstrap the semantics of existing tools
which might have an \emph{ad hoc} type system, such as the \lstinline|udis86|
disassembler for x86~\citep{udis86} or the ARM XML
specification~\citep{arm_xml_spec}.

\macaw{} was built to support an industry research team developing novel binary
analysis tools.
This team executed on multiple research projects over a decade, with a
significant amount of personnel change over this time.
In this environment, \macaw{} has served as a key library that helped us
develop successful tool prototypes.
In this paper, we discuss the most interesting and mature tool designs, which
are:
\begin{itemize}
\item \reopt{}, a binary lifter which converts \xEightySix{} binaries to LLVM
code and performs reoptimizations on the lifted code. \emph{This was the
original \macaw{} use-case.}
\item \saw, a formal verification platform that supports mixed C and machine
code. See~\citep{saw_s2n,Boston2021,saw_blst} for industry use-cases of SAW.
\item \renovate, a static binary rewriter that allows users to add or remove
code without executing the binary.
\item \mcTrace, a binary instrumentation tool to insert probes to dynamically
collect telemetry with low-overhead tracing, including in
environments without an operating system.
\item \ambient, a static verifier that can prove the presence of weird
machines~\citep{weirdMachines} in binaries (or lack thereof).
\item \pate, a relational verifier that proves that two binaries have the same
observable behaviors.
\item \cerridwen, a tool for ranking how similar a binary is to a known corpus
of binaries.
\item \surveyor, an interactive debugger that, among other features, can step
through machine code and inspect symbolic values.
\end{itemize}
To illustrate in more depth how \macaw{} works, we discuss two of these
tools in detail, \reopt{} and \saw{}.
We focus on these two because they support very different tasks and, as such,
require combining \macaw's components in very different and illustrative ways.

A key component of \macaw's design is its deep integration with
\crucible{}~\citep{crucible}, an SMT-backed symbolic execution engine.
Although \crucible{} was originally developed for simulating higher-level
imperative programming languages such as C, the same technology proves valuable
for simulating the behavior of machine code as well.
Many tools built on top of \macaw{}---including \saw, \renovate, \ambient,
\pate, and \surveyor---leverage \crucible{} to peform different analyses,
including static program analysis and formal verification.
Because \crucible{} also supports higher-level languages such as C, this makes
it easier to simultaneously analyze codebases that mix these languages with
machine code.
This is something that \saw{} takes advantage of, as it leverages \macaw{} and
\crucible{} to verify mixed C and machine code.

%\macaw{} provides a unique point in the design space of binary analysis
%frameworks.
%%
%We compare \macaw{} to existing frameworks such as Angr, and we also compare
%the components of \macaw's ecosystem to similar tools.
%%
%Many of the differences come down to particular design choices and engineering
%tradeoffs; we describe both the upsides and downsides of the choices that led
%to \macaw's design.

\macaw{} provides a specific point in the design space of binary analysis
frameworks which has proved useful for us in developing a wide range of
capabilities.
In \Cref{sec:macaw-design}, we discuss the core design of the \macaw{} IR and
supporting ecosystem, and we discuss how this design aligns with goals.
In \Cref{sec:case-studies}, we explore the tools that we have built on top of
\macaw{} in more detail.
We use this to explain how \macaw{} helps us to effectively build tools in a
variety of domains.
In \Cref{sec:related-work}, we compare \macaw{} to existing frameworks such as
Angr, and we also compare \macaw-based tools to similar efforts.
Many of the differences come down to tradeoffs; we describe both the upsides
and downsides of the choices that led to \macaw's design.
\Cref{sec:future-work} concludes and discusses our future design
ambitions for \macaw.

The source code for \macaw{} is publicly available at
\url{https://github.com/GaloisInc/macaw}.

\section{Design of the \macaw{} Ecosystem}\label{sec:macaw-design}

\usetikzlibrary{arrows.meta,positioning}

\begin{figure*}
  \begin{center}
  \resizebox{0.9\textwidth}{!}{%
  \begin{tikzpicture}[
              > = stealth, % arrow head style
              shorten > = 1pt, % don't touch arrow head to node
              auto,
              node distance = 2.5cm, % distance between nodes
              thick % line style
          ]
  \begin{scope}[transform shape]
    \node[nodeStyle] (macawBase) {\macawBase};
    \node[nodeStyle] (crucible) [below=0.6cm of macawBase, xshift=-3.0cm] {\crucible};
    \node[nodeStyle] (macawSymbolic) [below=0.6cm of macawBase] {\macawSymbolic};
    \node[nodeStyle] (macawRefinement) [below=0.6cm of macawSymbolic] {\macawRefinement};

    \node[draw=none] (macawDialectsText) [right=1.5cm of macawBase] {\footnotesize \macaw{} dialects};
    \node[nodeStyle] (macawAArch32) [below=0.05cm of macawDialectsText] {\macawAArch};
    \node[nodeStyle] (macawPPC) [below=0.15cm of macawAArch32] {\macawPPC};
    \node[nodeStyle] (macawRISCV) [below=0.15cm of macawPPC] {\macawRISCV};
    \node[nodeStyle] (macawX86) [below=0.15cm of macawRISCV] {\macawX};
    \node[draw,dotted,fit=(macawDialectsText) (macawAArch32) (macawPPC) (macawRISCV) (macawX86)] (macawDialects) {};

    \node[nodeStyle] (supportLibraries) [right=1cm of macawDialects] {\footnotesize ISA-specific libraries};

    \path[edgeStyle] (macawSymbolic) edge node {} (crucible);
    \path[edgeStyle] (macawSymbolic) edge node {} (crucible);
    \path[edgeStyle] (macawSymbolic) edge node {} (macawBase);
    \path[edgeStyle] (macawRefinement) edge node {} (crucible);
    \path[edgeStyle] (macawRefinement) edge node {} (macawSymbolic);
    \path[edgeStyle] (macawDialects) edge node {} (supportLibraries);

    % Hack to make the arrows horizontal:

    \draw[->] (2.2,-1.2) -- (1.5,-1.2); % Arrow between MacawDialects -> MacawSymbolic
    \draw[->] (2.2,0) -- (1.2,0); % Arrow between MacawDialects -> MacawBase
  \end{scope}
  \end{tikzpicture}%
  }

  \caption{The \macaw{} library ecosystem.
    An arrow A $\rightarrow$ B indicates that library A depends on library B.}

  \label{fig:macaw-libraries}

  \end{center}
 \end{figure*}

The \macaw{} ecosystem consists of a set of Haskell libraries, whose
relationships are depicted in \Cref{fig:macaw-libraries}.
At the heart of the ecosystem is \macawBase{}, which only contains
functionality that is independent of any particular architecture.
This includes the core \macaw{} intermediate representation (IR), code
discovery, optimization passes, and an ELF loader for ingesting binaries.

Any architecture-specific functionality is encapsulated in a \macaw{} dialect
library.
\macaw{} was originally designed with \xEightySix{} support in mind, which gave
rise to the first dialect, \macawX.
Later, \macawAArch, \macawPPC, and \macawRISCV{} were added.
Each of these dialects are built on top of other libraries that integrate
\macaw{} with disassemblers for specific instruction set architectures (ISAs),
and they also encode the semantics of each architecture's instructions into a
form that can be reasoned about.

% \todo{Say a bit more about the other topics covered in this section?}

\subsection{\macawBase: Core Data Types and Algorithms}

The \macawBase{} library is central to how \macaw{} works, as it defines the
key data types and operations needed to perform many binary analysis tasks.
All of \macaw's operations revolve around an intermediate language (defined in
\macawBase) where each IR program consists of a sequence of basic blocks.
This IR contains all of the control flow and ISA-independent functionality
needed to perform general analyses, but it is also designed to be extensible so
that ISA-specific functionality can be added in particular \macaw{} dialects.

We built \macawBase{} with statically typed functional programming techniques
in mind.
Various binary analysis-related algorithms can be expressed in relatively
succinct fashion by applying ideas from functional programming (see
\Cref{sec:mhnf}).
Moreover, we use \macaw{} is used in verification, so there needs to be
assurance the tooling and machine code semantics are correct.
To improve assurance, our implementation encodes information about register
sizes, operation bitwidths, and architectural aspects statically using
Haskell’s rich type system, thus statically checking many correctness
properties of the implementation.

\subsubsection{Typed Addresses}

A key data type that is used in almost every \macaw{} algorithm is its notion
of machine addresses.
\macawBase{} represents a machine address with the \lstinline|MemWord w| type,
which consists of a machine word of size \lstinline|w| bits:

\begin{lstlisting}[language=Haskell]
newtype MemWord (w :: Nat) = MemWord Word64
\end{lstlisting}

Note that \lstinline|w| is a \emph{type-level} number, which means that this
convention is checked at compile-time, not runtime.
This makes critical use of GHC's \lstinline|DataKinds| extension, which allows
natural numbers to be promoted to the type level using the \lstinline|Nat|
kind~\citep{dataKinds}.
For instance, 32-bit addresses have the type \lstinline|MemWord 32|, 64-bit
addresses have the type \lstinline|MemWord 64|, and it is a type error to mix
up the two types without performing an explicit conversion.

The \lstinline|w| type parameter is one example of a key design choice made
throughout \macaw's code: where possible, use GHC language extensions to encode
invariants about machine code at the type level.
This tradition was inspired by the \crucible{} symbolic execution
library~\citep{crucible}, with which \macaw{} integrates (see
\Cref{sec:symbolic-execution}).

One example of where this strong typing discipline is put to use is in how
\macaw{} encodes the bit widths of AArch32 instructions.
AArch32 processors have two instruction sets: ARM mode, where each instruction
is 32 bits, and Thumb mode, where each instruction is 16 bits.
\macaw{} encodes these bit widths at the type level, and as a result,
attempting to mix the two instructions sets without an explicit mode switch
will result in a type error.
This has caught many potential errors when developing \macaw-based tools, as
ARM and Thumb instructions can often occur within the same binary.

Note that although a \lstinline|MemWord| represents an address, it does not
encode any information about where the address is in a binary (aside from a raw
number), nor does it guarantee that the memory at that address will be valid.
\macawBase{} defines additional data types to attend to these needs.
At one level of abstraction higher than a \lstinline|MemWord| is a
\lstinline|MemAddr w|, which consists of a region number (\lstinline|addrBase|)
and an offset into the region (\lstinline|addrOffset|).
Note that multiple \lstinline|MemAddr|s can inhabit the same region:

\begin{lstlisting}[language=Haskell]
data MemAddr w = MemAddr
  { addrBase :: Int
  , addrOffset :: MemWord w
  }
\end{lstlisting}

A \lstinline|MemAddr| value represents a relocatable region which may (or may
not) be mapped to actual memory.
\macaw{} adopts the convention that the region number \lstinline|0| always
refers to absolute addresses.
Other region numbers are used to represent concepts such as
position-independent code.

When \macaw{} encodes a binary, it needs to know information about each
\emph{segment}, i.e., each logically distinct sequence of a in memory, along
with the corresponding addresses and permissions.
The \lstinline|MemAddr| type alone does not encode all this information, so
\macawBase{} contains additional abstractions for this.
First, \macawBase{} defines a \lstinline|MemSegment| data type, which describes
the overall contents of a single segment:

\begin{lstlisting}[language=Haskell]
data MemSegment w = MemSegment
  { segmentBase :: Int
  , segmentOffset :: MemWord w
  , segmentFlags :: Flags
  , segmentContents :: SegmentContents w
  }
\end{lstlisting}

Like a \lstinline|MemAddr|, a \lstinline|MemSegment| contains a region number
(\lstinline|segmentBase|) and an offset into that region
(\lstinline|segmentOffset|).
Unlike \lstinline|MemAddr|, a \lstinline|MemSegment| also stores whether the
segment has read, write, or execute permissions (\lstinline|segmentFlags|), and
it also maps each address in the segment to its underlying memory contents
(\lstinline|segmentContents|).
The \lstinline|SegmentContents| data type describes the mapping of addresses to
memory:

\begin{lstlisting}[language=Haskell]
newtype SegmentContents w =
  SegmentContents (Map (MemWord w) (MemChunk w))

data MemChunk w
  = ByteRegion ByteString
  | RelocationRegion (Relocation w)
  | BSSRegion (MemWord w)

data Relocation w = ...
\end{lstlisting}

The \lstinline|MemChunk| type characterizes whether a chunk of memory is a
sequence of specific bytes (\lstinline|ByteRegion|), a relocatable region whose
contents are computed by a \lstinline|Relocation| value
(\lstinline|RelocationRegion|), or a \lstinline|.bss| section that is
initialized with all zeroes (\lstinline|BSSRegion|).
The precise details of how different parts of binaries map to which
\lstinline|MemChunk|s are beyond the scope of this paper.

\macawBase{} uses the \lstinline|MemSegmentOff| data type to describe an
address that is guaranteed to be valid (i.e., point to valid memory).
A \lstinline|MemSegmentOff| value is simply a \lstinline|MemSegment| plus an
offset into the that segment:

\begin{lstlisting}[language=Haskell]
data MemSegmentOff w = MemSegmentOff
  { segoffSegment :: MemSegment w
  , segoffOffset :: MemWord w
  }
\end{lstlisting}

Finally, a binary is a collection of \lstinline|MemSegment|s, along with
metadata describing what the entrypoint address is and which region numbers map
to which segments.

% \macaw{} resolves a \lstinline|MemAddr w| value into a
% \lstinline|MemSegmentOff w| value.
% %
% Unlike \lstinline|MemAddr|s, a \lstinline|MemSegmentOff| is guaranteed to be a
% statically valid address.
% %
% In turn, a \lstinline|Memory w| value contains a collection of
% \lstinline|MemSegmentOff| values that consist of all the statically known
% contents of memory.

% At the next level of abstraction is \lstinline|MemAddr w|, which consists of a
% region number and an offset into the region.
% %
% A \lstinline|MemAddr| value represents a relocatable region which may (or may
% not) be mapped to actual memory.
% %
% \macaw{} adopts the convention that the region number \lstinline|0| always
% refers to absolute addresses.
% %
% Other region numbers are used to represent concepts such as
% position-independent code.
%
% \macaw{} resolves a \lstinline|MemAddr w| value into a
% \lstinline|MemSegmentOff w| value.
% %
% Unlike \lstinline|MemAddr|s, a \lstinline|MemSegmentOff| is guaranteed to be a
% statically valid address.
% %
% In turn, a \lstinline|Memory w| value contains a collection of
% \lstinline|MemSegmentOff| values that consist of all the statically known
% contents of memory.

\subsubsection{Typed Memory Shapes}

Another place where \macaw{} draws inspiration from \crucible{} is in how
\macaw{} encodes the \emph{shapes} of values that are stored in memory.
Although all data in a binary can be thought of as a series of bytes, the shape
of those bytes is very much dependent on the context in which it is used.
For example, some machine instructions interpret bytes as integers, some
instructions interpret bytes as floating-point values, and so on.
It is all too easy to mix up different shapes, so \macaw{} encodes these shapes
at the type level and parameterizes different operations based on what shapes
they expect.

Specifically, \macawBase{} defines a \lstinline|MacawType| data type, which is
shown in~\Cref{fig:macaw-type}.
Note that the data constructors of \lstinline|MacawType| are only ever meant to
be used at the type level (again, using GHC's \lstinline|DataKinds| extension),
which gives \macaw{} uses a pseudo-dependently typed flair.
Aside from encoding invariants about memory shapes at the type level, it is
also useful to be able to check what sort of \lstinline|MacawType| one has at
runtime.
This is achieved via a \lstinline|TypeRepr| object, which acts as a
\emph{singleton type} for \lstinline|MacawType|~\citep{singletons}.
One can pattern-match on a \lstinline|TypeRepr| value to determine which sort
of \lstinline|MacawType| is used at runtime.

\begin{figure*}[h]
\begin{tabular}{ll}
  \begin{lstlisting}[language=Haskell, boxpos=t]
  data MacawType
    = -- | A bitvector
      BVType Nat
      -- | Floating-point
    | FloatType FloatInfo
      -- | A vector of types
    | VecType Nat Type

  data FloatInfo
    = -- | 32-bit IEEE754
      SingleFloat
    | -- | 64-bit IEEE754
      DoubleFloat
    | ...
  \end{lstlisting}
  &
  \begin{lstlisting}[language=Haskell, boxpos=t]
  data TypeRepr (tp :: MacawType) where
    BVTypeRepr
      :: NatRepr n
      -> TypeRepr (BVType n)
    FloatTypeRepr
      :: FloatInfoRepr fi
      -> TypeRepr (FloatType fi)
    VecTypeRepr
      :: NatRepr n
      -> TypeRepr tp
      -> TypeRepr (VecType n tp)

  data FloatInfoRepr (fi :: FloatInfo) where
    SingleFloatRepr ::
      FloatInfoRepr SingleFloat
    DoubleFloatRepr ::
      FloatInfoRepr DoubleFloat
    ...
  \end{lstlisting}
  \end{tabular}
  \caption{The definition of \lstinline|MacawType| in \macaw{}, which describes the
           shape of machine-code values at the type level. This type is
           witnessed at the value level by \lstinline|TypeRepr|, which can be
           thought of as the singleton type~\citep{singletons} for
           \lstinline|MacawType|.}
  \label{fig:macaw-type}
\end{figure*}

\subsubsection{From Addresses to Architectures}%
\label{sec-addrs-to-archs}

Just as \macaw's address types are indexed by the number of bits
\lstinline|(w :: Nat)| or a memory shape \lstinline|(tp :: MacawType)|, many \macaw{}
abstract syntax types are parameterized by an \lstinline|arch| type parameter,
which represents the processor architecture being used.
Doing so ensures that operations from separate architectures are not mixed up.
For example, \macawX{} instantiates \lstinline|arch| with an \lstinline|X86_64|
type, \macawAArch{} instantiates \lstinline|arch| with an \lstinline|AArch32|
type, and so on.
We refer to types like \lstinline|X86_64| and \lstinline|AArch32| as
\emph{architecture extension points}.
One example of an \lstinline|arch|-indexed type is \lstinline|ArchReg|, which
is displayed in more detail in~\Cref{fig:arch}.
\lstinline|ArchReg| uses GHC's \lstinline|TypeFamilies| extension
~\citep{openTypeFunctions} to map an architecture extension point to a register
data type.

\begin{figure*}
  %\begin{minipage}{0.49\textwidth}
  \begin{lstlisting}[language=Haskell]
  type family ArchReg arch :: MacawType -> Type
  type instance ArchReg X86_64 = X86Reg
  type instance ArchReg AArch32 = ARMReg
  ...

  type family RegAddrWidth (r :: MacawType -> Type) :: Nat
  type instance RegAddrWidth X86Reg = 64
  type instance RegAddrWidth ARMReg = 32
  ...

  type family ArchStmt arch :: MacawType -> Type
  type ArchAddrWidth arch = RegAddrWidth (ArchReg arch)
  type ArchAddrWord arch = MemWord (ArchAddrWidth arch)
  type ArchMemAddr arch = MemAddr (ArchAddrWidth arch)
  type ArchSegmentOff arch = MemSegmentOff (ArchAddrWidth arch)
  \end{lstlisting}
  %\end{minipage}
  \caption{Type families and synonyms related to architecture extension points.}
  \label{fig:arch}
\end{figure*}

Sometimes, \lstinline|arch|-indexed data types must interact with
\lstinline|w|-indexed data types.
A \macaw{} architecure uniquely determines the number of bits in a machine
address, so \macaw{} defines a \lstinline|RegAddrWidth| type family to bridge
the gap between the former and the latter.
This is such a commonly used type family that \macaw{} also provides a variety
of type synonyms defined on top of \lstinline|RegAddrWidth| for convenience,
which are depicted in~\Cref{fig:arch}.

\subsubsection{The \macaw{} Intermediate Language}

\begin{figure*}
\begin{tabular}{ll}
  \begin{lstlisting}[language=Haskell, boxpos=t]
  data Block arch ids = Block
    { blockStmts ::
        [Stmt arch ids]
    , blockTerm ::
        TermStmt arch ids }

  data TermStmt arch ids
    = FetchAndExecute
        RegState
         (ArchReg arch)
         (Value arch ids)
    | TranslateError
        RegState
         (ArchReg arch)
         (Value arch ids)
        Text
    | ArchTermStmt
        ArchTermStmt arch
         (Value arch ids)
        RegState
         (ArchReg arch)
         (Value arch ids)

  type family ArchStmt arch
    :: (MacawType -> Type) -> Type
  type family ArchTermStmt arch
    :: (MacawType -> Type) -> Type
  \end{lstlisting}
  &
  \begin{lstlisting}[language=Haskell, boxpos=t]
  data Stmt arch ids where
    AssignStmt
      :: Assignment arch ids tp
      -> Stmt arch ids
    WriteMem
      :: ArchAddrValue arch ids
      -> MemRepr tp
      -> Value arch ids tp
      -> Stmt arch ids
    ExecArchStmt
      :: ArchStmt arch
          (Value arch ids)
      -> Stmt arch ids
    ArchState
      :: ArchMemAddr arch
      -> RegState
          (ArchReg arch)
          (Value arch ids)
      -> Stmt arch ids
    InstructionStart
      :: ArchAddrWord arch
      -> Text
      -> Stmt arch ids
    Comment
      :: Text
      -> Stmt arch ids
  \end{lstlisting}
  \end{tabular}
  \caption{Core data types used in \macaw's intermediate language.}
  \label{fig:block}
\end{figure*}

\macaw{} uses a three address code--based intermediate
language~\citep{dragonBook} that is
centered around basic blocks (\lstinline|Block|).
Each basic block consists of a list of statements (\lstinline|Stmt|) followed
by a special terminator statement (\lstinline|TermStmt|).
These data types are given in~\Cref{fig:block}.

Each data type is parameterized by two type parameters: \lstinline|arch|, which
encodes the architecture, and \lstinline|ids|, which denotes the set of
identifiers used in assignment instructions.
The \lstinline|ids| type parameter is unique to each decoded function, which
ensures that instructions from different functions cannot be mixed without an
explicit renaming step.
The \lstinline|ids| type parameter functions similarly to the \lstinline|s|
type parameter in Haskell's \lstinline|ST s| type~\citep{launchbury1995state}.

The control flow in \macaw{} is oriented around \lstinline|Block|s, and a
disassembled machine instruction can correspond to multiple \lstinline|Stmt|s
in a \lstinline|Block|, depending on the complexity of the instruction.
Note that the \lstinline|Stmt|s in two blocks are allowed to overlap.
For example, this is necessary to support AArch32 binaries, where instructions
can be in both ARM or Thumb mode depending on how they are reached.

A single \lstinline|Stmt| can perform an assignment (\lstinline|AssignStmt|),
write to memory (\lstinline|WriteMem|), execute an ISA-specific statement
(\lstinline|ExecArchStmt|), or update register values (\lstinline|ArchState|).
The \lstinline|Stmt| type is also used to encode metadata about which parts of
a basic block correspond to the start of an instruction
(\lstinline|InstructionStart|) and any additional information that is useful
for debugging purposes (\lstinline|Comment|).
The \lstinline|Stmt| data type has many auxiliary types used in its data
constructors, so we will define ones that are relevant to this paper as they
become relevant.

% A \lstinline|Stmt| can assign a virtual register (\lstinline|AssignStmt|),
% write to memory (\lstinline|WriteMem|), update a machine register
% (\lstinline|ArchState|), or invoke an architecture-specific extension
% (\lstinline|ExecArchStmt|).
% %
% Note that \macaw{} makes a distinction between virtual registers and machine
% registers.
% %
% There are unlimited virtual registers, with each intermediate value given a
% distinct virtual register, much like LLVM's virtual registers.
% %
% There are a finite number of machine registers, which correspond to the
% registers used in a particular architecture (e.g., \lstinline|rax| in
% \xEightySix).
%
% The most common form of \lstinline|TermStmt| is \lstinline|FetchAndExecute|,
% which stands for the fetching and execution of the upcoming instruction in the
% processor state.
% %
% This may be known to be the instruction at a concrete address, such as the next
% instruction, when a block simply falls through to a subsequent block, or the
% statically known target of a direct jump or call.
% %
% But the value of the instruction pointer may also be a complex symbolic
% computation, for instance, the result of a computed jump or an indirect call
% that may involve symbolic inputs or symbolic values read from memory.
% %
% There is also an \lstinline|ArchTermStmt| constructor for architecture-specific
% terminal statements.
% %
% If a block ends prematurely due to an error in instruction decoding, the block
% is terminated with \lstinline|TranslationError|, which contains the register
% state at the time of the error so that \macaw{} can attempt to resume execution
% from that point.

The core \macaw{} IR is compact and covers common operations present in
multiple ISAs, which are used to represent the behavior of many instructions.
These operations primarily involve integer arithmetic (e.g., \lstinline|add|
and \lstinline|sub| in \xEightySix), and \macaw{} encodes these operations
using bitvectors in a way that can be easily translated to SMT.
More complicated operations (i.e., ones that do not map directly to SMT) are
put into syntactically separate parts of the \lstinline|Stmt| language.
For example, memory accesses are their own category, as they must be decomposed
into SMT-friendly operations via symbolic execution (see
\Cref{sec:symbolic-execution}).
There is also a separate category for architecture-specific extensions (e.g.,
data cache hints in PowerPC), for which each ISA has a small number of
instructions.
Depending on where these instructions can occur in a basic block, these
ISA-specific instructions are put into \lstinline|ArchStmt| or
\lstinline|ArchTermStmt|.

Separating common operations from more complicated ones reflects a key \macaw{}
design decision: the \macaw{} IR does not attempt to be generic enough to
encode every possible instruction directly.
One benefit for this decision is that it keeps the \macaw{} IR relatively
readable.
Although \macaw{} \emph{could} expand every instruction directly to core
operations, this could easily cause the number of \lstinline|Stmt|s to balloon
in size.
For example, \xEightySix's AES-NI instructions would likely need hundreds of
bitvector operations to encode directly.
Another benefit is that some instructions perform effects that not all \macaw{}
clients need to model, so putting them into architecture-specific extensions
means that clients only ``pay'' for extensions that they reason about.
This gives \macaw{} users a high degree of control when mixing different
components.

\subsubsection{Code Discovery}\label{sec:code-discovery}

% One of the most important passes in \macawBase{} is code discovery, which aims
% to identify all of the functions in a binary that are reachable from a set of
% entry point addresses.
% %
% \macaw's code discovery algorithm was built with the following design goals
% in mind:
%
% \paragraph{Human readability} The results of code discovery should be simple
% enough that it can be checked against an ISA manual.
% %
% It should also be straightforward to trace IR terms to individual instructions.
%
% \paragraph{Graceful error recovery} Code discovery should never throw fatal
% exceptions.
% %
% Instead, discovery failures should be explicitly represented in the IR and
% allow the algorithm to continue as much as possible.
%
% \paragraph{Minimal requirements} The discovery algorithm should be able to produce
% results even if a binary is not compiled with symbols or debug information.
% %
% That is, providing extra symbols may allow the algorithm discover \emph{more}
% things, but it should not be a hard requirement.
%
% \paragraph{Semantic control flow} The control flow of a program should be
% identified semantically, rather than relying on \macaw{} to pattern match
% against particular sequences of instructions.
%
% \paragraph{Incrementality} It should be possible to discover a basic blocks in
% isolation, separately from discovering the rest of the binary.
% %
% This is important both for performance and for being able to compute partial
% results.

\begin{figure*}
  \begin{center}
  \resizebox{0.9\textwidth}{!}{%
  \begin{tikzpicture}[
              > = stealth, % arrow head style
              shorten > = 1pt, % don't touch arrow head to node
              auto,
              node distance = 2.5cm, % distance between nodes
              thick % line style
          ]
  \begin{scope}[transform shape]
    \node[nodeStyle, draw opacity=0] (entryPoints) {\textit{Entry point}\\\textit{addresses}};
    \node[nodeStyle] (explorationFrontier) [right=1.5cm of entryPoints] {Exploration frontier};
    \node[nodeStyle] (classify) [below=0.9cm of explorationFrontier, xshift=-2cm] {Classify};
    \node[nodeStyle] (disassembleBlock) [right=2cm of classify] {Disassemble block};
    \node[nodeStyle] (rewrite) [below=2.5cm of explorationFrontier] {Rewrite (simplify)};
    \node[nodeStyle, draw opacity=0] (discoveredFunctions) [left=0.5cm of classify, yshift=-1.6cm] {\textit{Discovered} \\ \textit{functions}};

    \path[edgeStyle] (entryPoints) edge node {} (explorationFrontier);
    \path[edgeStyle] (explorationFrontier) edge node {} (disassembleBlock);
    \path[edgeStyle] (disassembleBlock) edge node {} (rewrite);
    \path[edgeStyle] (rewrite) edge node {} (classify);
    \path[edgeStyle] (classify) edge node {\textit{Branch}} (disassembleBlock);
    \path[edgeStyle] (classify) edge node {\textit{Call}} (explorationFrontier);
    \path[edgeStyle] (classify) edge node {\textit{Return}} (discoveredFunctions);

  \end{scope}
  \end{tikzpicture}%
  }

  \caption{\macaw's core code discovery algorithm.}

  \label{fig:code-discovery}

  \end{center}
 \end{figure*}

One of the most important passes in \macawBase{} is code discovery.
The core discovery pass identifies all of the functions in a binary that are
reachable from a set of entry point addresses.
This pass also distinguishes intra-procedural control flow like branches,
jumps, and switch statements from inter-procedural control flow such as calls,
tail calls, and returns.
Code discovery is required for compositional code analysis passes and designed
to work even when a binary lacks symbol tables or other static mechanisms for
identifying layout.

The core algorithm that powers code discovery is illustrated
in~\Cref{fig:code-discovery}.
The algorithm discovers one function at a time by taking candidate function
addresses from the \emph{exploration frontier}, which is seeded with an initial
set of entry points.
Usually, these entry points are taken from the entry point address of an ELF
binary, the dynamic entry points of a shared library, the addresses of static
symbols (when available), as well as function addresses derivable from
metadata, such as DWARF unwinding tables.
\macaw{} also supports specifying custom entry points, which can be useful for
binaries that lack symbol information (e.g., stripped binaries).

The main part of the algorithm is the block decoding loop, which discovers
functions by decoding their individual blocks.
First, \macaw{} decodes an instruction using a low-level disassembly function
for the specific architecure.
Next, \macaw{} lifts the instruction into a semantic representation as a
sequence of \lstinline|Stmt|s.
If the instruction changes control flow (e.g., \xEightySix's \lstinline|call|),
\macaw{} will terminate the block with a \lstinline|TermStmt|; otherwise, it
will restart the loop and continue decoding the block.

Next, \macaw{} performs rewriting, which simplifies terms in the block to
improve readability and simplify later analyses.
Afterwards, \macaw{} analyzes the block to determine what kind of control-flow
transfer terminates the block, be it a jump, function call, or
otherwise.
To this end, \macaw{} tracks abstract domains for each machine register, using
techniques inspired by value-set analysis~\citep{valueSetAnalysis} to update the
domains as new blocks are discovered.
At the end of analysis, \macaw{} consults the abstract return address, as well
as the addresses and bounds of jump tables, to \emph{classify} how the block
terminates.
Targets of function calls are added to the exploration frontier as they are
classified, which is used in subsequent iterations of the algorithm.

\paragraph{An example rewriting pass}\label{sec:mhnf}

One optimization that \macaw's rewriting pass enables is converting values to
\emph{mux head normal form} (MHNF), where if-then-else expressions (i.e.,
muxes) are raised as far up in the CFG as possible.
As an example, consider this minimized CFG (pretty-printed for compactness):

\begin{lstlisting}
r1 := ...
r2 := Mux r1 0x4 0x28
r3 := Add r2 0x100a3e50
{ ip => r3 }
\end{lstlisting}

This CFG is not in MHNF, as it contains a mux expression (\lstinline|r2|) that
is nested within another expression's definition (\lstinline|r3|).
To convert this CFG to MHNF, we want to rewrite it into something that looks
more like this:

\begin{lstlisting}
r1 := ...
r4 := Mux r1 0x100a3e54 0x100a3e78
{ ip => r4 }
\end{lstlisting}

Where \lstinline|0x100a3e54| and \lstinline|0x100a3e78| have folded away the
constant addition of \lstinline|0x100a3e50| to \lstinline|0x4| and
\lstinline|0x28|, respectively.
This way, the instruction pointer (\lstinline|ip|) value is directly a mux
expression.
MHNF CFGs are preferable from a code discovery perspective, as \macaw's code
discovery pass is more likely to notice branch statements when all of the
branching is encoded as a top-level mux in the instruction pointer register.

One nice consequence of \macaw{} being written in a functional language is that
the MHNF pass can be implemented in a very simple, easy-to-follow, and
relatively terse fashion.
The MNHF pass (as well as the rest of the rewriter's passes) are implemented in
the \lstinline|Rewriter| monad, which is a state monad that carries around the
necessary state as well as the list of statements that have been rewritten so
far:

\begin{lstlisting}
newtype Rewriter arch s src tgt a =
  Rewriter (StateT (RewriteState arch s src tgt) (ST s) a)

data RewriteState arch s src tgt =
  RewriteState
    (RewriteContext arch s src tgt)
    [Stmt arch tgt]
\end{lstlisting}

Most of the details of \lstinline|RewriteContext| are not essential to
understanding the MNHF pass, so we elide them here.
A useful helper function that we need in order to simplify constant expressions
(such as the \lstinline|Add| expression in the example above) is
\lstinline|rewriteApp|:

\begin{lstlisting}
rewriteApp ::
  App (Value arch tgt) tp ->
  Rewriter arch s src tgt (Value arch tgt tp)
rewriteApp app =
  case app of
    BVAdd _ x (BVValue _ 0) -> do
      pure x
    BVAdd w (BVValue _ x) (BVValue _ y) -> do
      pure (BVValue w (toUnsigned w (x + y)))

    -- No normal rewrites available,
    -- now try MHNF for enabling code discovery
    _ -> rewriteMhnf app
\end{lstlisting}

The full definition of \lstinline|rewriteApp| is omitted for brevity, but the
code above shows the essence of the function.
Expressions that can be simplified with simple syntactic rewrites are
simplified, and everything else is delegated to \lstinline|rewriteMhnf|, the
function which performs the MHNF pass:

\begin{lstlisting}
rewriteMhnf ::
  App (Value arch tgt) tp ->
  Rewriter arch s src tgt (Value arch tgt tp)
rewriteMhnf app =
  case app of

    BVAdd w
        (valueAsApp -> Just (Mux p c t f))
        v@(BVValue _ _) -> do
      t' <- rewriteApp (BVAdd w t v)
      f' <- rewriteApp (BVAdd w f v)
      rewriteApp $ Mux p c t' f'

    BVAdd w
        (valueAsApp -> Just (Mux p c t f))
        v@(RelocatableValue _ _) -> do
      t' <- rewriteApp (BVAdd w t v)
      f' <- rewriteApp (BVAdd w f v)
      rewriteApp $ Mux p c t' f'

    -- no more rewrites applicable, so return the final result
    _ -> evalRewrittenRhs (EvalApp app)

-- Add an assignment statement that evaluates the right hand side
-- and return the resulting value.
evalRewrittenRhs ::
  AssignRhs arch (Value arch tgt) tp ->
  Rewriter arch s src tgt (Value arch tgt tp)
\end{lstlisting}

This function recognizes if-expressions (i.e., \lstinline|Mux|es) that appear
nested underneath bitvector additions (i.e., \lstinline|BVAdd|s)
If it encounters such an if-expression, it pushes the addition down through the
branches of the if-expression, simplifying constant expressions with
\lstinline|rewriteApp| along the way.
If it encounters any other type of expression, then it assigns the expression
to a fresh identifier, records the assignment in the \lstinline|RewriteState|,
and returns the assigned expression as a \lstinline|Value| via
\lstinline|evalRewrittenRhs|.

The MHNF rewriting pass involves a number of distinct steps: constant
simplification, pattern recognition, the MNHF rewrites themselves, and
appending new assignments to the current state.
Thanks to the use of the \lstinline|Rewriter| monad, we are able to separate
the concerns of appending new assignments from the ``business logic'' of
performing rewrites.
Thanks to Haskell's pattern matching, recognizing the shapes of if-expressions
is straightforward.
And thanks to classic functional programming techniques, we are able to
compartmentalize the tedious (but mechanical) details of simplifying constant
expressions in a single \lstinline|rewriteApp| function.

% \begin{figure*}
%  \begin{center}
%  \begin{tikzpicture}[
%              > = stealth, % arrow head style
%              shorten > = 1pt, % don't touch arrow head to node
%              auto,
%              node distance = 2.5cm, % distance between nodes
%              thick % line style
%          ]
%  \begin{scope}[transform shape]
%   \node[nodeStyle] (entryPoints) {Entry point \\ addresses};
%   % \node[nodeStyle] (explorationFrontier) [right=1cm of entryPoints] {Exploration frontier};
%   % \node[nodeStyle] (disassembleBlock) [right=1cm of explorationFrontier] {Disassmble block};
%   % \node[nodeStyle] (rewrite) [below=1cm of disassembleBlock] {Rewrite \\ (simplify)};
%   % \node[nodeStyle] (classify) [below=1cm of rewrite] {Classify};
%   % \node[nodeStyle] (discoveredFunctions) [below=1cm of classify] {Discovered \\ functions};
%
%   % \path[edgeStyle] (entryPoints) edge node {} (explorationFrontier);
%   % \path[edgeStyle] (explorationFrontier) edge node {} (disassembleBlock);
%   % \path[edgeStyle] (disassembleBlock) edge node {} (rewrite);
%   % \path[edgeStyle] (rewrite) edge node {} (classify);
%   % \path[edgeStyle] (classify) edge[bend right,right] node {Branch} (disassembleBlock);
%   % \path[edgeStyle] (classify) edge[bend left] node {Call} (explorationFrontier);
%   % \path[edgeStyle] (classify) edge node {Return} (discoveredFunctions);
%  \end{scope}
%  \end{tikzpicture}
%
%  \caption{\macaw's core code discovery algorithm.}
%
%  \label{fig:code-discovery}
%
%  \end{center}
% \end{figure*}

\subsubsection{Symbolic Execution via \crucible}\label{sec:symbolic-execution}

A key component of the \macaw{} ecosystem is \macawSymbolic{}, which defines
functionality for simulating \macaw{} IR programs with \crucible{}, a symbolic
execution library targeting imperative code~\citep{crucible}.
Using \macawSymbolic{}, one can simulate a binary with symbolic data, such as
symbolic register states or stack contents.
During simulation, \crucible{} will generate verification conditions that are
discharged to an SMT solver.

Besides machine code, \crucible{} also includes dialects targeting other
imperative programming languages, including C, Java, and Rust.
\crucible's LLVM dialect (which enables reasoning about C code) has been
battle-tested on a large number of verification problems from the SV-COMP
verification competition~\citep{cruxSVComp}.
Moreover, \crucible{} can reason about code that mixes languages, such as C
programs that use inline assembly (see \Cref{sec:saw}).

\paragraph{Symbolic Code Discovery}\label{sec:macaw-refinement}

The \macaw{} ecosystem also includes an additional, optional method of code
discovery provided by the \macawRefinement{} library.
This library utilizes symbolic execution to discover code targets, which can
supplement the simpler, pattern-based heuristics that \macawBase{} uses.

As an example, consider a \lstinline|switch| statement whose expression is an
integer \lstinline|i| whose range of possible values was previously constrained
by an \lstinline|if| statement.
When the various case targets of the \lstinline|switch| statement are small and
regularly sized, the compiler may encode the \lstinline|switch| statement as a
jump to \lstinline|i * s|, where \lstinline|s| is the size of the largest case
block.
\macawBase's pattern-based heuristics will not be aware the constraints from
the \lstinline|if| statement, and thus they cannot limit the scope of possible
jump targets to the specific set of valid addresses.
Using \macawRefinement{}, on the other hand, allows an SMT solver to indicate
the constrained set of possible jump targets to add those to the discovery
frontier.
\subsection{\macawX: The First \macaw{} Dialect}

\begin{figure*}
\begin{center}
  % \resizebox{0.9\textwidth}{!}{%
  \begin{tikzpicture}[
              > = stealth, % arrow head style
              shorten > = 1pt, % don't touch arrow head to node
              auto,
              node distance = 2.5cm, % distance between nodes
              thick % line style
          ]
  \begin{scope}[transform shape]
    \node[nodeStyle] (binaryX86) {x86-64 binary};
    \node[nodeStyle] (flexdis86) [right=0.6cm of binaryX86] {Flexdis86\\(AST of disassembled\\instructions)};
    \node[nodeStyle] (macawX86) [right=0.6cm of flexdis86] {Macaw-X86\\ (Macaw CFG)};

    \path[edgeStyle] (binaryX86) edge node {} (flexdis86);
    \path[edgeStyle] (flexdis86) edge node {} (macawX86);
  \end{scope}
  \end{tikzpicture}%
  % }
\end{center}
\caption{The architecture of \macawX, describing the process of going from an
\xEightySix{} binary to a \macaw{} CFG.}
\label{fig:macaw-x86-64}
\end{figure*}

\macawX{}, which targets the x86-64 ISA, was the first ISA dialect that we
developed alongside the \reopt{} tool (see \Cref{sec:reopt}).
The architecture of \macawX{} is pictured in \Cref{fig:macaw-x86-64}.
While this figure is specific to \xEightySix, it establishes a general pattern
that we continued to use when developing future dialects, albeit with some
notable differences (which we will address later).

Because \macawX{} was our first attempt at a dialect, we did not have an
existing methodology for integrating an ISA into \macaw{}, nor did we have a
clear picture of what design choices one could make in the process of
developing a dialect.
One notable decision that we made (perhaps unconsciously) during development
was to hand-write most of the code in \macawX.
This is because the \reopt{} tool needed to be able to run on \xEightySix{}
binaries with a relatively quick turnaround time, and as such, we could not
wait until we had 100\% coverage of the \xEightySix{} ISA before demonstrating
results.
As such, we implemented support for individual \xEightySix{} instructions on an
as-needed basis, manually implementing whatever scaffolding was necessary for
\macaw{} to reason about each instruction.

\subsubsection{Disassembling \xEightySix{} Binaries Using \flexdis}

The first step that \macawX{} performs when analyzing a binary is to
\emph{diassemble} it---that is, to ingest the binary's raw bytes into a more
structured representation.
A binary's contents can be thought of as a stream of instructions, and the role
of a disassembler is to parse one instruction from the stream at a time.
We wrote our own \xEightySix{} disassembler library\footnote{\flexdis{}
contains both an \xEightySix{} disassembler and assembler, but we do not use
the assembler from \macaw.}, named \flexdis{}, which represents each
instruction as a Haskell data type, called an \lstinline|InstructionInstance|.
An \lstinline|InstructionInstance| captures all of an instruction's relevant
properties, such as its name, opcode, size, and operands (be they immediate
values, registers, addresses, or otherwise).

The main entrypoint into \flexdis{} is the \lstinline|disassembleInstruction|
function:

\begin{lstlisting}[language=Haskell]
disassembleInstruction ::
  ByteReader m => NextOpcodeTable -> m InstructionInstance
\end{lstlisting}

The \lstinline|ByteReader| class characterizes Haskell monads that can
represent instruction streams.
\flexdis{} is primarily meant to be used with a \lstinline|ByteReader| instance
that efficiently reads a binary from disk, but \flexdis{} also supports a
simpler \lstinline|ByteReader| instance that can be used to mock the contents
of a binary for testing purposes.

Because \xEightySix{} is a very large ISA with well over 1,000 instructions,
writing a feature-complete disassembler for \xEightySix{} is no small feat.
To make this task more manageable, \lstinline|disassembleFunction| takes an
\lstinline|NextOpcodeTable| argument, which represents a lookup table that maps
an instruction opcode to possible \lstinline|InstructionInstance| values:

\begin{lstlisting}[language=Haskell]
type NextOpcodeTable = Vector OpcodeTable

data OpcodeTable
   = -- There are still more opcode bytes to parse.
     OpcodeTable NextOpcodeTable
     -- All of the opcode bytes have been parsed, resulting in
     -- a list of possible instruction candidates.
   | OpcodeTableEntry !InstructionCandidate
\end{lstlisting}

To understand how \lstinline|NextOpcodeTable| works, consider the
\lstinline|pop| instruction.
Some versions of the \lstinline|pop| instruction have opcodes that begin with
the byte \lstinline|0f|, so a \lstinline|NextOpcodeTable| would have an entry
at index \lstinline|0f| containing a \lstinline|OpcodeTable| that handles the
remaining opcode bytes.
This \lstinline|OpcodeTable| would then contain another, smaller
\lstinline|NextOpcodeTable| with entries at indices \lstinline|a1| and
\lstinline|a9|, each containing \lstinline|OpcodeTableEntry| values for
\lstinline|pop|.
These table entries represent two variants of the instruction with the opcodes
\lstinline|0f a1| and \lstinline|0f a9|, respectively.
The smaller \lstinline|NextOpcodeTable| would also map any opcode bytes for
other instructions whose opcodes begin with \lstinline|0f| (e.g., the
\lstinline|addps| instruction, which can use the opcode \lstinline|0f 58|).

Constructing a \lstinline|NextOpcodeTable| is by far the most tedious part of
\flexdis's disassembler.
To make this process less tedious, we partially automate it by leveraging the
\lstinline|udis86| project~\citep{udis86}, which implements an x86 and
\xEightySix{} disassembler for C.
\flexdis{} uses \lstinline|udis86|'s XML table describing all \xEightySix{}
instructions to automatically generate a \lstinline|NextOpcodeTable| value.
This automation does save us quite a bit of time, but it is not perfect.
\flexdis's version of the XML table makes some manual tweaks, such as adding
additional instructions not covered by \lstinline|udis86|.
We also do not automate any of the logic for turning
\lstinline|InstructionCandidate|s into full-fledged
\lstinline|InstructionInstance|s, which can be surprisingly complicated due to
the sheer number of instruction prefixes that \xEightySix{} uses to distinguish
different instructions.

\subsubsection{Encoding \xEightySix{} Semantics Using \macawX}

After disassembling a binary's instructions, the next step is to reify them
into a semantic representation that \macaw{} can reason about.
The \macawX{} dialect accomplishes this by defining an
\lstinline|InstructionDef| for each \xEightySix{} instruction, which converts
an \lstinline|InstructionInstance| from \flexdis{} into a list of \macaw{}
basic blocks:
%
% \todo{We are referencing \macaw-specific definitions here, which makes me
% wonder if this section should actually come after the section that discusses
% \macawBase's data types.}

\begin{lstlisting}[language=Haskell]
newtype InstructionSemantics = InstructionSemantics
  (forall st ids. InstructionInstance -> X86Generator st ids ())

-- A monad for generating Macaw basic blocks.
data X86Generator st ids a

-- A pair of an instruction mnemonic (e.g., `pop`)
-- alongside its semantics.
type InstructionDef = (ByteString, InstructionSemantics)
\end{lstlisting}

We also define a small domain-specific language for defining
\lstinline|InstructionDef| values that makes it more obvious what an
instruction's semantics mean at a glance.
As a small example, here is the complete semantics for the \lstinline|cbw|
(convert byte to word) instruction in \macawX:

\begin{lstlisting}[language=Haskell]
def_cbw :: InstructionDef
def_cbw = defNullary "cbw" $ do
  v <- get al
  ax .= sext n16 v
\end{lstlisting}

These semantics use \lstinline|defNullary| to signify that the \lstinline|cbw|
instruction does not take any operands.
It first reads a byte from the \lstinline|al| register (\lstinline|get al|),
uses sign extension (\lstinline|sext|) to convert the byte into a two-byte
value (\lstinline|n16|), and finally copies the two-byte value into the
\lstinline|ax| register (\lstinline|ax .= ...|).
The \lstinline|InstructionDef| DSL also includes shorthand for other common
machine code idioms (immediate values, memory locations, offsets, etc.) that
are not shown above.

Unlike \flexdis, where we made an effort to automate at least some parts of the
implementation, all of the semantics implemented in \macawX{} are completely
hand-written.
This was largely a practical consideration, as we wanted to demonstrate results
for the subset of \xEightySix{} instructions that we encountered the most on
the \reopt{} project, and hand-writing the semantics for these hand-picked
instructions proved a fast way to accomplish this.
Our methodology was to carefully read the Intel \xEightySix{} ISA
manual~\citep{intelISA} and extract formal semantics for each instruction's
entry in the manual.
Although the list of instructions covered by \macawX's semantics has grown over
the years, it is still far from complete.

The choice to hand-write \macawX's semantics, while expedient in the
short-term, poses a challenge to long-term maintenance.
If we use \macawX{} on a binary containing an instruction that is not covered
by the list of covered instructions, then we must add the instruction's
semantics manually, which is a time-consuming process.
Not only that, but this is an \emph{error-prone} process, as human programmers
are liable to make mistakes when translating the ISA manual's prose.

\subsection{\macawPPC: A More Automated Dialect}

\begin{figure*}
\begin{center}
  \resizebox{0.9\textwidth}{!}{%
  \begin{tikzpicture}[
              > = stealth, % arrow head style
              shorten > = 1pt, % don't touch arrow head to node
              auto,
              node distance = 2.5cm, % distance between nodes
              thick % line style
          ]
  \begin{scope}[transform shape]
    \node[nodeStyle] (binaryPPC) {PowerPC binary};
    \node[nodeStyle] (dismantlePPC) [right=1.2cm of binaryPPC] {Dismantle-PPC\\(AST of disassembled\\instructions)};
    \node[nodeStyle] (semMCPPC) [right=1.2cm of dismantlePPC] {semMC-PPC\\ (What4 program)};
    \node[nodeStyle] (MacawPPC) [right=1.2cm of semMCPPC] {Macaw-PPC\\(Macaw CFG)};

    \draw[edgeStyle] (binaryPPC) -- (dismantlePPC) node[below, pos=0.5, yshift=-0.9cm] {(via LLVM\\TableGen)};
    \draw[edgeStyle] (dismantlePPC) -- (semMCPPC) node[below, pos=0.5, yshift=-0.9cm]
            {(via hand-written\\base set semantics +\\automated SemMC\\semantics translation\\via Crucible)};
    \draw[edgeStyle] (semMCPPC) -- (MacawPPC)node[below, pos=0.5, yshift=-0.9cm] {(via Macaw-SemMC)};

  \end{scope}
  \end{tikzpicture}%
  }
\end{center}
\caption{The architecture of \macawPPC, describing the process of going from an
PowerPC binary to a \macaw{} CFG.}
\label{fig:macaw-ppc}
\end{figure*}

As we gained experience developing the \macawX{} dialect, it became
increasingly clear that manually curating dialects for large ISAs did not scale
well as well as we'd like.
We decided that if we were to add another \macaw{} dialect for a big ISA, we
would make a concerted effort to automate as much of the process as was
reasonable.
That opportunity came when we developed the next dialect, \macawPPC.
The architecture of \macawPPC{} is pictured in \Cref{fig:macaw-ppc}.
Our focus with \macawPPC{} was to support both 32- and 64-bit PowerPC binaries,
although we tend to exercise the 32-bit PowerPC code paths more often due to
its prevalence in embedded software.

\subsubsection{Disassembling PowerPC Binaries Using \dismantle}

The first step in developing \macawPPC{} was to generate a disassembler.
We used the \lstinline|udis86| disassembler in order to (partially) generate
\macawX's disassembler, but because \lstinline|udis86| is specific is x86, we
could not do something similar PowerPC.
As a result, we developed a more general solution: a cross-architecture
assembler and diassembler generation library called \dismantle.
\dismantle{} would become the basis for \macawPPC's disassembler, as well as
the the disassembler for most \macaw{} dialects going forward.

\dismantle{} works by consulting the data provided by LLVM's TableGen
tool~\citep{llvmTablegen}.
LLVM uses TableGen as part of its code generator to describe the encodings,
types, and operands of machine instructions.
TableGen data is also ideal for generating disassemblers, so \dismantle{} is
able to use TableGen with little or no changes required.\footnote{Typically,
the only time where we need to modify the TableGen data for \dismantle's
purposes is to work around LLVM bugs.}

TableGen data files are large, and we don't want to have to load a data file
every time we disassemble a PowerPC binary, as this would be prohibitively
expensive.
Our solution is to instead load the data file only once when compiling the
package.
We use Template Haskell~\citep{templateHaskell} to load the PowerPC data file
at compile time, parse it, and then use Template Haskell to generate code for
an efficient disassembler based on the TableGen information.
Note that we trade runtime efficiency in exchange for much more work at
compile-time: it can take several minutes to compile all of \dismantle{}.
This is a theme that would continue to reappear as we moved in the direction of
increased automation.

\subsubsection{Semi-Automated PowerPC Semantics Using \semmc}

Just like with disassembling PowerPC, we sought to futher automate the process
of encoding the semantics of PowerPC instructions into \macaw.
Luckily for us, \citet{stratifiedSynthesis} had recently published work on how
to automatically generate formal semantics for \xEightySix using a techique
called \emph{stratified synthesis}.
Although this work came out well after we hand-wrote our own formal semantics
for \xEightySix, we decided to apply stratified synthesis when developing
PowerPC semantics.
We developed a library for performing this form of synthesis (for PowerPC and
for other ISAs) and named it \semmc{} (\textsc{Sem}antics of \textsc{M}achine
\textsc{C}ode).

The \semmc{} library is designed to learn machine code semantics by starting
with a core \emph{base set} of instructions for which the semantics are
hand-written.
Using this base set, we execute \semmc{} on a series of fuzzer-generated
machine states and instructions and use the results to learn the semantics of
instructions that are not in the base set.
To further increase our confidence that the learned semantics are correct, we
run the randomly generated machine states on actual PowerPC hardware and use
that as an oracle for what behavior should be considered correct.
We repeat this process until we have a complete formal description of the ISA's
semantics.
In this sense, \semmc's stratified synthesis is semi-automated: some manual
effort is required to describe the semantics of the base set, but everything
else is fully automated.

Much like with the semantics implementation in \macawX, the semantics of
\semmc{} base set instructions are specified using a DSL.
For example, here are the semantics of PowerPC's \lstinline|nand| instruction
in \semmc:

\begin{lstlisting}[language=Haskell]
definePPCOpcode NAND xform3c $ \rA rB rS -> do
  comment "NAND (X-form)"
  let res = bvnot (bvand (Loc rS) (Loc rB))
  defLoc rA res
\end{lstlisting}

Here, \lstinline|definePPCOpcode| takes the instruction name and what
instruction form to use.\footnote{Instruction forms describe how instructions
are encoded in PowerPC. It is not essential to understand how instruction forms
work to read this section.}
We use \lstinline|xform3c|, which indicates an X-form instruction with three
operands.
These operands are then bound as \lstinline|rA|, \lstinline|rB|, and
\lstinline|rC|, which can then be used in the instruction's definition.
Finally, we use \lstinline|defLoc| to store the result of performing a
bitwise-AND of the values from the \lstinline|rS| and \lstinline|rB| operands
into the location that the operand \lstinline|rA| uses.

\semmc{} converts the DSL into an intermediate language called \whatFour.
\whatFour{} programs are stored on disk using an S-expression--based
representation so that upstream programs (e.g., \macawPPC) can parse and
interpret the semantics they contain.
For example, the \whatFour{} representation of the \lstinline|nand| semantics
above is the following S-expression:

\begin{lstlisting}[language=Lisp]
;; NAND (X-form)
((operands
 ((rA Gprc)
  (rB Gprc)
  (rS Gprc)))
 (in
  (op.rS op.rB loc.IP))
 (defs
  ((op.rA
   (with
    ()
    (bvnot
     (bvand op.rS op.rB))))
   (loc.IP
    (with
     ()
     (bvadd loc.IP #x0000000000000004))))))
\end{lstlisting}

The \whatFour{} representation of \semmc{} semantics explicitly indicate which
operands are inputs or outputs (via the \lstinline|in| section), and it
precisely captures how each output is modified when invoking the instruction.
In addition to updating the \lstinline|rA| operand, the \whatFour{} version
also updates the instruction pointer \lstinline|IP|, which is a lower-level
detail that was elided in the DSL encoding.

The DSL is a convenient way for humans to write \whatFour{} semantics in the
base set, but \semmc{} also needs to be able to synthesize \whatFour{}
semantics based on fuzzing.
This is a more complicated process, as \semmc{} must be able to reason about
how a machine's state updates when invoking an instruction, and it must be able
to do this where not all parts of the machine state are known ahead of
time---that is, when some parts of the machine state are \emph{symbolic}.

To make this process simpler, \semmc{} leverages the
\crucible{}~\citep{crucible} symbolic execution library.
\crucible{} is primarily designed for imperative programming languages, such as
C or Rust, but \crucible{} also works well for machine code as well.
\semmc{} leverages \crucible{} when synthesizing \whatFour{} semantics for
instructions not in the base set.

Finally, once we have \whatFour{} semantics for all PowerPC instructions, we
translate the \whatFour{} programs into Macaw CFGs using the \macawSemMC{}
library.
As was the case with \dismantle, we do not want to perform all of this
conversion at runtime, as we would need to read many, many What4 S-expression
files in order to look up the semantics for all of the instructions used in a
typical PowerPC program.
As such, \macawSemMC{} uses Template Haskell to look up all What4 S-expression
files at compile time and generate Macaw CFG--based semantics for each PowerPC
instruction, thereby avoiding the need to consult any What4 files at runtime.

The payoff of introducing \semmc{} as part of the semantics translation is that
we gain a significant amount of automation in the overall process, thereby
avoiding some of the maintenance headaches that pervade \macawX's fully manual
approach to semantics.
There is a cost to doing so, however: \semmc's extra layers of abstraction make
it so that it approximately two minutes to compile the semantics for the entire
PowerPC ISA.

\subsection{\macawAArch: An Even More Automated Dialect}

\begin{figure*}
\begin{center}
  % \resizebox{0.9\textwidth}{!}{%
  \begin{tikzpicture}[
    >=Stealth,
    shorten >=1pt,
    node distance=2cm,
    thick
  ]

  %--- Nodes ---
  \node[nodeStyle] (aarch32) {AArch32 binary};

  \node[nodeStyle, right=3.5cm of aarch32] (dismantle)
    {Dismantle-ARM-XML\\(AST of dismantled instructions)};

  \node[nodeStyle, below=2cm of aarch32] (asl)
    {ASL-Translator\\(Crucible CFG)};

  \node[nodeStyle, right=of asl] (semMC)
    {SemMC-AArch32\\(What4 program)};

  \node[nodeStyle, right=2.7cm of semMC] (macaw)
    {Macaw-AArch32\\(Macaw CFG)};

  %--- Edges ---
  \path[edgeStyle]
    (aarch32)
      edge node[above] {via official ARM\\XML specification}
      (dismantle);

  \path[edgeStyle]
    (dismantle)
      edge node[above,sloped] {via official ARM\\XML specification}
      (asl);

  \path[edgeStyle]
    (asl)
      edge node[above] {via Crucible}
      (semMC);

  \path[edgeStyle]
    (semMC)
      edge node[above] {via \\Macaw-SemMC}
      (macaw);

\end{tikzpicture}
% }
\end{center}
\caption{The architecture of \macawAArch, describing the process of going from an
AArch32 binary to a \macaw{} CFG.}
\label{fig:macaw-aarch32}
\end{figure*}

Using \semmc{} to automate the development of \macawPPC{} offered significant
advantages in how confident we were in the resulting PowerPC semantics'
correctness, and it made the code more maintainable as well.
The \semmc{} approach is not foolproof, however.
\semmc{} still requires human beings to curate semantics for base set
instructions by hand, and while the size of the base set is smaller than the
size of the overall ISA, curating the base set can still be a tedious (and
potentially error-prone) process.

When developing \macawAArch, a \macaw{} dialect for the 32-bit ARM ISA, we had
an opportunity to develop the level of automation even further.
ARM maintains machine-readable, executable specifications of every
instruction~\citep{arm_xml_spec}, which gives \macawAArch{} an official source
of truth for how to disassemble and interpret any given instruction.
Combining the offical ARM specifications with Template Haskell allows us to
automate almost every aspect of \macawAArch's development.

The architecture of \macawAArch{} is pictured in \Cref{fig:macaw-aarch32}.
Like \macawPPC, \macawAArch also relies on the \dismantle{} package in order to
disassemble binaries.
One notable difference in how \dismantle{}'s PowerPC and AArch32 dialects work
is that \dismantle{} uses LLVM TableGen data for PowerPC, whereas it uses the
ARM XML specifications for AArch32.
Note that the constructing parse tables via the XML specifications is quite
slow: it takes over an hour to build tables for the entire ISA.
As such, we pre-compute the parse tables and store them on disk so that they
can simply be loaded directly during subsequent rebuilds of \dismantle.

There is also a \semmc{} dialect for AArch32, but unlike the \semmc{} dialect
for PowerPC, it does not generate semantics for ARM instructions using
stratified synthesis.
Instead, it uses a dedicated library (named \aslTranslator) to convert the ARM
XML specification's semantics directly into the \whatFour{} intermediate
language that \semmc{} uses.
The XML specification is written in ASL~\citep{arm_xml_spec}, a custom-purpose
imperative programming language with a strong types.
As it turns out, the typing discipline that ASL uses is a natural fit for
\crucible, so \aslTranslator heavily relies on \crucible{} to translate the
imperative, ASL-based semantics into a purely functional, \whatFour-based
encoding.

Note that \aslTranslator{} is almost, but not completely, automated.
\aslTranslator{} manually defines a handful of primitives to give semantics to
low-level ASL operations (e.g., bitvector slicing, querying the program counter
value, etc.)
This is by far the trickiest part of \aslTranslator{} to write, and the vast
majority of bug reports ultimately step from incorrectly implementing these
primitives.

Like \macawPPC{}, \macawAArch's heavy use of Template Haskell (both in
\aslTranslator and in other parts of the dialect) incurs a significant
compile-time performance penalty.
In fact, the performance penalty is even more severe for \macawAArch: it takes
well over 5 minutes to compile the parts of the dialect that rely on
\aslTranslator-generated semantics.
While it is likely that these times could be improved by working harder to
reduce the size of the generated code, this reflects the general trend that the
more one uses Template Haskell, the longer one has to wait.

\subsection{\macawRISCV: Revisiting a Manually-Written Dialect}

\begin{figure*}
\begin{center}
  % \resizebox{0.9\textwidth}{!}{%
  \begin{tikzpicture}[
    > = stealth,         % arrow head style
    shorten > = 1pt,     % don't touch arrow head to node
    auto,
    node distance = 2.5cm, % distance between nodes
    thick               % line style
  ]
  \begin{scope}[transform shape]
  % First node: RISC-V binary
  \node[nodeStyle] (binaryRiscv) {RISC-V binary};

  % Second node: GRIFT (AST of disassembled instructions)
  \node[nodeStyle] (grift) [right=0.6cm of binaryRiscv]
  {GRIFT\\(AST of disassembled\\instructions)};

  % Third node: Macaw-RISCV (Macaw CFG)
  \node[nodeStyle] (macawRiscv) [right=0.6cm of grift]
  {Macaw-RISCV\\(Macaw CFG)};

  % Edges (arrows)
  \path[edgeStyle] (binaryRiscv) edge node {} (grift);
  \path[edgeStyle] (grift) edge node {} (macawRiscv);
  \end{scope}
\end{tikzpicture}
% }
\end{center}
\caption{The architecture of \macawRISCV, describing the process of going from an
RISC-V binary to a \macaw{} CFG.}
\label{fig:macaw-riscv}
\end{figure*}

The most recently introduced \macaw{} dialect is \macawRISCV, which targets the
RISC-V architecture.
The architecture of \macawRISCV{} is pictured in \Cref{fig:macaw-riscv}.
In stark contrast to previous \macaw{} dialects, which have moved increasingly
in the direction of automatically generating disassemblers and semantics for
their respective ISAs, \macawRISCV{} implements its RISC-V disassembler and
semantics completely by hand.
This choice was motivated by the fact that RISC-V adheres to the reduced
instruction set computer (RISC) philosophy.
The base set of RISC-V instructions (not including extensions) only consists of
48 instructions, which are composed of a relatively small number of different
opcodes.
As such, it is viable to hand-write a \macaw{} dialect for RISC-V.

\subsection{The Manual-Automated Spectrum}

\begin{figure*}
  \begin{center}
  \begin{tikzpicture}[>=Stealth]

    % Draw the main spectrum arrow (left to right)
    \draw[<->, thick] (0,0) -- (8,0)
        node[right]{Fully automated};

    % Left label
    \node[left] at (0,0) {Fully manual};

    % Title above the arrow
    \node[above] at (4,0.3) {\textit{Manual--Automated Spectrum}};

    % Helper macro: "tick" for vertical dashes
    % usage: \tick{x}{heightAbove}{heightBelow}
    \newcommand{\tick}[3]{%
        \draw[thick] (#1,#2) -- (#1,#3);
    }

    % Place ticks and labels for each category
    % You can adjust x-coordinates to space them out as you wish
    % and the "tick" heights for the dash length.
    \tick{1}{0.15}{-0.15}
    \node[below, rotate=-20, anchor=north west] at (1,-0.2) {Macaw-RISCV};

    \tick{3}{0.15}{-0.15}
    \node[below, rotate=-20, anchor=north west] at (3,-0.2) {Macaw-x86};

    \tick{5}{0.15}{-0.15}
    \node[below, rotate=-20, anchor=north west] at (5,-0.2) {Macaw-PPC};

    \tick{7}{0.15}{-0.15}
    \node[below, rotate=-20, anchor=north west] at (7,-0.2) {Macaw-AArch32};

\end{tikzpicture}
\end{center}
\caption{How much automation each \macaw{} dialect requires to implement on a
scale from ``Fully manual'' (entirely written by humans) to ``Fully automated''
(derived entirely from external sources by machines).}
\label{fig:manual-automated-spectrum}
\end{figure*}

Each \macaw{} dialect exists on a spectrum of how much automation is used to
implement the code supporting the dialect's ISA, as depicted in
\Cref{fig:manual-automated-spectrum}.
One far end of the spectrum is hand-writing all of the code (e.g.,
\macawRISCV).
The opposite end of the spectrum is fully automating the generation of all
ISA-related code (no dialect meets this criterion exactly, but \macawAArch{}
comes the closest).

Having developed four \macaw{} dialects, we have honed some intuition regarding
whether to use a hand-written approach, an automated approach, or somewhere in
between when creating a new dialect.
There are various questions one must ask to determine which approach is right
for a given ISA:

\begin{itemize}

 \item How many different instructions are there?
 It is possible to exhaustively go through all of the instructions in a small,
 RISC architecture, but for non-RISC architectures, the pain of going through
 every instruction will be more pronounced.

 \item How clear is the ISA specification regarding instruction semantics?
 Different ISA semantics are written with different levels of formality.
 For instance, the ARM XML specification is written with extreme formality,
 while the Intel \xEightySix{} ISA manual is less so, sometimes requiring
 humans to interpret dense English-language prose in order to determine how
 certain instructions behave.
 For the latter case, a \semmc-like approach can help.

 \item How tolerant are you of long compile times?
 While automatically generating code with Template Haskell can increase ISA
 coverage and confidence in the generated code, it comes with a severe downside
 of noticeably increasing compile times.
 It is worth asking the question of whether this compile-time slowdown is
 acceptable for a given project.

\end{itemize}

\section{Case Studies}\label{sec:case-studies}

\subsection{\reopt}\label{sec:reopt}

\begin{figure*}
  \begin{center}
    \begin{tikzpicture}[
                > = stealth, % arrow head style
                shorten > = 1pt, % don't touch arrow head to node
                auto,
                node distance = 2.5cm, % distance between nodes
                thick % line style
            ]
    \begin{scope}[transform shape]
      \node[nodeStyle,inner sep=0,draw opacity=0] (entryPoints) {\textit{Initial entry}\\\textit{points}};

      \node[nodeStyle] (blockDiscovery) [right=0.7cm of entryPoints] {Incremental\\block discovery};
      \node[nodeStyle] (blockRecovery) [above=0.8cm of blockDiscovery, xshift=1.5cm] {Block recovery};
      \node[draw=none] (macawPhaseText) [above=-0.5cm of blockRecovery, xshift=-2.2cm] {\macaw{}};
      \node[draw,inner sep=2mm,dotted,fit=(blockDiscovery) (blockRecovery) (macawPhaseText) ] (macawPhase) {};

      \node[draw=none] (reoptPhaseText) [right=7.5cm of macawPhaseText] {\reopt{}};
      \node[nodeStyle] (constraintGeneration) [left=1.5cm of reoptPhaseText, yshift=-0.4cm] {Constraint\\generation};
      \node[nodeStyle] (constraintSolving) [below=0.5cm of constraintGeneration] {Constraint\\solving};
      \node[nodeStyle] (functionRecovery) [below=0.5cm of constraintSolving] {Func recovery};
      \node[nodeStyle] (llvmGeneration) [right=0.75cm of functionRecovery] {LLVM gen};
      \node[draw,inner sep=2.5mm,dotted,fit=(constraintGeneration) (constraintSolving) (functionRecovery) (llvmGeneration) (reoptPhaseText) ] (macawPhase) {};

      \node[nodeStyle, draw opacity=0] (outputNode) [right=0.7cm of llvmGeneration] {};

      \path[edgeStyle] (entryPoints) edge node {} (blockDiscovery);
      \path[edgeStyle] (blockDiscovery) edge node {} (blockRecovery);
      \path[edgeStyle] (blockRecovery) edge node {} (constraintGeneration);
      \path[edgeStyle] (constraintGeneration) edge node {} (constraintSolving);
      \path[edgeStyle] (constraintSolving) edge node {} (functionRecovery);
      \path[edgeStyle] (functionRecovery) edge node {} (llvmGeneration);
      \path[edgeStyle] (llvmGeneration) edge node {} (outputNode);

      \path[edgeStyle] (functionRecovery) edge node {\textit{New entry points}\\ \textit{identified}} (blockDiscovery);
    \end{scope}
    \end{tikzpicture}
   \caption{\reopt's function recovery loop.}

 \label{fig:reopt-function-recovery}
 \end{center}
\end{figure*}

\reopt~\citep{hendrix2019towards} is an optimization tool that lifts functions in
a compiled binary into LLVM bitcode, optimizes the intermediate bitcode, and
then recompiles it into a new executable.
\reopt{} currently supports \xEightySix{} binaries compiled in the ELF format
for Linux.
The \macaw{} library began its life as an internal component of the \reopt{}
tool, and much of \macaw's design reflects considerations that were important
at the time of \reopt's development.
The source code for \reopt{} is publicly available at
\url{https://github.com/GaloisInc/reopt}.

The primary challenge in designing a tool like \reopt{} is going from
\macaw{}'s IR, which only has very minimal type information associated with
each basic block, to the LLVM IR, which has a comparatively richer type system.
At the \macaw{} level, most values in general-purpose registers look like
integers, but when \reopt{} outputs LLVM bitcode, it is helpful to distinguish
between integers, structs, pointers, and function pointers.

To this end, \reopt{} includes a type-based function recovery pass, which is
depicted in \Cref{fig:reopt-function-recovery}.
Starting from a list of initial entry point functions (which are either
inferred from the binary or user-specified), the tool calls \macaw{} to
discover the closure of basic blocks reachable from the currently known blocks.
Once the basic blocks have been recovered (c.f.
~\Cref{sec:code-discovery}), \reopt{} performs a demand analysis,
allowing it to compute the \xEightySix{} arguments and return registers on a
per-block basis, aggregating them into register ``demand sets'' for entire
functions, which are then synthesized into appropriate argument and return
values for the whole function.

At this point, we only know the size of registers, but not whether they contain
integers, floating-point values, or, for those registers whose size matches that
of virtual addresses for the architecture, data or code pointers.
However, there is information in the instruction flow that can help distinguish
between all these cases, sometimes uniquely identifying what high-level type a
register should have.
In order to produce better LLVM code, \reopt{} will gather constraints for all
intermediate registers, function arguments, and return values.  For instance, a
register flowing into the memory operand of a memory operation ought to be some
pointer, while a value coming out of a division operation must be numeric.
\reopt{} will solve these constraints to determine a candidate type for every
function.  \reopt{} also has some experimental support for row types, allowing
it to recover structural types, including recursive ones such as linked lists.

During this process, \reopt{} may uncover calls to other functions, in which
case \macaw's code discovery will need to be invoked once more.  For instance,
if a function's argument is determined to hold function pointers, all concrete
values flowing into this argument position can be safely added to the candidate
entry points pool.  Thanks to \macaw{}'s incremental design, \reopt{} is able to
resume block discovery with those new entry points without incurring duplicate
work.
Once all of the constraints are solved, the final output is an LLVM program
consisting of all the function definitions that have been recovered, which now
have much richer types.

\subsubsection{\reopt{} Type Constraints}

\reopt's approach to reverse engineering types for assembly functions is loosely
based on the \TIE{} system described in~\citet{Lee2011}.
The type system presented in that work is slightly more sophisticated than what
\reopt{} needs.
For instance, \reopt{} generates LLVM, which does not distguish between the
types of signed and unsigned integers, so \reopt{} does not bother to track
this information.
Likewise, where \TIE{} relies on value-set analysis to get more precise offset
information in structured types, \reopt{} relies on \macaw{}'s symbolic offsets,
which may either be concrete offsets or opaque to its analysis.

Ultimately, \reopt{} is concerned with distinguishing pointers from non-pointer
values.
Without the type inference pass, \reopt{} would lift all pointer
arithmetic to integer arithmetic and would have to introduce an LLVM
\lstinline|inttoptr| cast every time such a value would be in address position
in a memory \lstinline|load| or \lstinline|store| instruction.
This is important because treating pointers as integers up to the last second
inhibits LLVM optimizations and obfuscates pointer manipulation for both human
and tool consumers of the lifted IR.
Instead, pointer arithmetic can be lifted to pointer-manipulating instructions,
describing them using types with richer structure, benefiting the readability of
the lifted code, and enabling more complex code analyses and transformations.
Knowing these types, \reopt{} can produce the more appropriate
\lstinline|getelementptr| LLVM instruction, operating over typed pointers.
We give an instance of such an output change
in~\Cref{fig:getelementptr}.

\begin{figure*}[h]
  \begin{tabular}{ll}
  \begin{lstlisting}[language={[x86masm]Assembler}]
;; Original x86-64 instruction
cmpb   $0x30,0x3(%rax)           ; in AT&T syntax
cmp    BYTE PTR [rax+0x3],0x30   ; in Intel syntax

;; Translation without type inference
;; %t0 corresponds to %rax and is given type i64
%t1 = add i64 %t0, 0x3
%t2 = load i8, i8* inttoptr (i64 %t1 to i8*)
%t3 = icmp eq i8 %t2, 0x30

;; Translation with type inference
;; %t0 corresponds to %rax and is given type {{i8,i8,i8},{i8,i8}}*
%t1 = getelementptr {{i8,i8,i8},{i8,i8}},
                    {{i8,i8,i8},{i8,i8}}* %t0, i32 0, i32 1
%t2 = getelementptr {i8,i8}, {i8,i8}* %t1, i32 0, i32 0
%t3 = load i8, i8* %t2
%t4 = icmp eq i8 %t3, 0x30
  \end{lstlisting}
  \end{tabular}
  \caption{Example of the impact of \reopt~'s type inference on LLVM output.
  \lstinline|getelementptr| instructions only help LLVM's type system and have no performance
  impact.}
  \label{fig:getelementptr}
\end{figure*}

\subsubsection{\reopt{} Limitations}

\reopt's function recovery, as sophisticated as it is, may still not be enough
to infer a precise type for a function.
One reason is that \macaw's code discovery is not perfect, and when it cannot
process all of the instructions in a basic block, it will conservatively bail
out and possibly omit some instructions that are crucial for inferring the
types of the block's inputs and outputs.
To help alleviate this problem, \reopt{} includes an option to supply header
files, as well as debug information files, which contain function type
signatures that \reopt{} should assume when it encounters known functions.
This is particularly useful for external function calls (e.g., from
\lstinline|libc|), where \reopt{} either does not have the entire function
definition available, or the full definition is sufficiently low-level that
analyzing it proves costly.

Recent work has also been aimed at addressing limitations when working with
stripped binaries with no debug sections.
Such binaries have historically proven challenging for \reopt{}, as the only
entry point initially known for such binaries is the one listed as entry point
of the executable.
Typically, this will be the address of a symbol that would have been named
\lstinline|_start| before stripping, and whose code does \emph{not} contain an
explicit call to the address of the \lstinline|main| symbol.

\begin{lstlisting}[language={C}]
void _start() {
  // Some scaffolding code, ending with:
  __libc_start_main(&main); // `main` called back when C runtime ready
}
\end{lstlisting}

Instead, the start routine will call the C run-time scaffolding symbol,
\lstinline|__libc_start_main|, passing the address of the \lstinline|main|
symbol as an argument to it.
As such, the binary does not contain any explicit call to \lstinline|main|, and
\macaw{} will subsequently miss the \lstinline|main| function and stop exploring
right away.
This is where the type-driven analysis of \reopt{} can help.
In this case, \reopt{} knows that \lstinline|__libc_start_main| expects a
function pointer thanks to the \lstinline|libc| debug information.
Upon noticing the address of \lstinline|main| flowing into a call to
\lstinline|__libc_start_main|, \reopt{} will instruct \macaw{} to resume block
discovery with this new address as a function entry point.

There are three reasons why we designed \reopt{} to use a type-based approach.
First, \reopt{} favors this approach over heuristical ones, as this not only applies
to discovering the \lstinline|main| symbol in a stripped binary, but also any
function address passed to a higher-order function, be it in an external library
or locally within the binary being lifted.
Second, \reopt{} prefers a discovery-guided approach, where we only inspect
bytes of the code section when we witness an explicit flow of control to it,
over a greedy inspection of the entire code section, as it still allows us to
avoid false positives such as dead code, nop paddings, or misaligned decodings.
Finally, \reopt{} favors a type-aware approach to a greedy inspection of
all pointer-sized values that could look like they could point to code, as some
floating-point values may also look like pointers to the code segment, and this
would also yield false positives.
In all the above cases, not only does \reopt{} find known-to-be-reachable
addresses, it also discovers them in a context where it knows the type of their
arguments and return value, which allows \reopt{} to properly lift these basic
blocks into well-typed LLVM functions.

\subsection{\saw}\label{sec:saw}

The Software Analysis Workbench (\saw)~\citep{saw} is a verification tool that
can reason about the behavior of imperative programs by proving them equivalent
to functional specifications.
\saw{} supports a variety of different source languages, including code written
in C, Java, Rust, and most relevantly for this work, \xEightySix.
\saw{} has been successfully used to verify a variety of industrial-strength cryptographic
libraries written in a mixture of C and \xEightySix, including the s2n
Transport Layer Security (TLS) stack~\citep{saw_s2n}, the AWS LibCrypto
library~\citep{Boston2021}, and the BLST library~\citep{saw_blst}.
The source code for \saw{} is publicly available at
\url{https://github.com/GaloisInc/saw-script}.

\saw{} ingests \xEightySix{} machine code by leveraging \macaw{} to lift
assembly instructions into a \crucible{} control-flow graph.
Separately from the machine code itself, \saw{} also takes as input a list of
specifications that describe the intended behavior of each assembly function.
These specifications consist of Hoare-style pre- and post-conditions that
describe the shape of memory and the behavior of the function itself.
\saw{} checks each function against its specification by symbolically executing
the function's CFG and ensuring that the conditions described in the function
are respected along the way, discharging any non-trivial proof goals to an SMT
solver.

Aside from checking functional correctness, \saw{} also checks that a function
is memory safe.
In the context of a language like C, this is a well understood problem, as the
C standard dictates what it means for a C program to interact with memory in a
valid way.
The notion of memory safety becomes blurrier with assembly code, however, as
there are almost no universally accepted rules on what constitutes safe reads
and writes.
The only guidelines that one can typically rely on are the conventions used
within a particular program, as well as the Application Binary Interface (ABI)
for calling external functions.

% Nearly all \saw{} use cases for machine code verification involve mixed C and
% \xEightySix{} code, where \xEightySix{} functions are defined in a way that
% follow C-like conventions and respect the ABI.
% %
% As a result, \saw{} leverages \crucible's LLVM-based memory model as a basis
% for modeling \xEightySix.
% %
% There incurs the risk that some valid assembly programs will be rejected under
% the rules of the LLVM memory model.
% %
% In practice, however, nearly all of the machine code that we have verified with
% \saw{} is also safe under the LLVM memory model.

\saw{} specifically targets programs with mixed C and \xEightySix{} code where
the C code controls the program's memory, and all \xEightySix{} functions only
ever access the C code's memory.
In this setting, \saw{} can leverage \crucible's LLVM-based memory model for
modeling both types of code.
This means that \saw{} is able to prove all memory reads and writes in the
\xEightySix{} code are correct with respect to the host language's memory
model.
As a small example of how SAW's \xEightySix{} support works, we will verify the
behavior of a tiny \lstinline|increment| assembly function that takes a single
pointer argument and increments the value it points to by one.
\begin{center}
\begin{tabular}[t]{ll}
\begin{lstlisting}[language={[x86masm]Assembler}]
increment:
  mov rax, [rdi]
  add rax, 1
  mov [rdi], rax
  ret
\end{lstlisting}
&\hspace{2.5cm}
\begin{lstlisting}[language=C]
// Equivalent C code:
void increment(uint64_t* i)
{
  *i += 1;
}
\end{lstlisting}
\end{tabular}
\end{center}

We can write a SAW specification \lstinline|increment_spec| that captures the
intended behavior:

\begin{center}
\begin{tabular}[t]{l}
\begin{lstlisting}
let increment_spec = do {
  ptr <- llvm_alloc (llvm_int 64);
  val <- llvm_fresh_var "val" (llvm_int 64);
  llvm_points_to ptr (llvm_term val);
  llvm_execute_func [ptr];
  llvm_points_to ptr (llvm_term {{ val + 1 }});
};
\end{lstlisting}
\end{tabular}
\end{center}

This specification declares a series of statements, where each statement before
the \lstinline|llvm_execute_func| is a pre-condition and each statement after
is a post-condition.
The pre-conditions state that the function accepts a single pointer argument
that points to an arbitrary 64-bit LLVM integer value.
The post-condition states that after the function is invoked, the pointer
argument will then point to the initial integer value plus one.
Again, we are leveraging LLVM to describe the shape of memory, and all of the
commands used here could apply to C functions just as much as \xEightySix{} ones.

The \lstinline|increment| function can be checked against
\lstinline|increment_spec| by running:

\begin{center}
\begin{tabular}[t]{l}
\begin{lstlisting}
llvm_verify_x86 "./increment_bin" "increment" increment_spec z3;
\end{lstlisting}
\end{tabular}
\end{center}

Here, \lstinline|increment_bin| is the ELF binary that defines
\lstinline|increment|, and \lstinline|z3| is the SMT solver to discharge proof
goals to during symbolic execution.
This is the only command that is machine code--specific, as this is where
\macaw{} is used to lift the assembly code in \lstinline|increment| to a CFG
and pass it on to \saw's symbolic execution engine.
Although the \lstinline|llvm_verify_x86| command is specific to the
\xEightySix{} ISA, it could very easily be extended to handle other
architectures as well ~\footnote{For example, an
\lstinline|llvm_verify_aarch32| command is planned:
% \url{https://github.com/GaloisInc/saw-script/pull/987}
\emph{anonymized for review}
}.

\subsection{Other Case Studies}

% \macaw{} has been used as the basis for a variety of different binary analysis
% tools, which we briefly review here.

\paragraph{\ambient}

A verification tool built on top of \macaw{} that evaluates
whether or not it is possible to trigger a \emph{weird machine} (WM)
~\citep{weirdMachines} in a binary, i.e., a code execution path that occurs
outside of the intended specification of the program.
One challenging aspect of triaging the presence of WMs is determining what
environmental conditions---such as command-line arguments, environment
variables, or file system state---are strictly required to trigger the WM, and
which conditions are irrelevant to the WM.
To this end, \ambient{} leverages symbolic execution to make certain parts of
the environment symbolic to explore the different execution paths, which can
help users of the tool determine which parts of the environment are most
relevant to the WM at hand.

Many WMs reside in a shared library that a binary depends on, such as
\lstinline|libc|.
To this end, it is important to be able symbolically execute the code inside of
low-level shared libraries, but this can be difficult due to how often these
libraries contain low-level functions and system calls.
While \crucible{} can simulate this type of code, the performance is often
impractically slow.

To allow the tool to execute shared libraries in an efficient manner,
\ambient{} includes a mechanism for overriding the behavior of functions and
system calls.
For instance, \ambient{} can intercept calls to the \lstinline|getppid| system
call and replace it with a version that returns a fresh, symbolic integer as
the process ID.
While this does not faithfully emulate the actual implementation of
\lstinline|getppid|, this can be an acceptable compromise to make to speed up
symbolic execution, especially for WMs where the behavior of
\lstinline|getppid| is irrelevant.

The source code for \ambient{} is publicly available at
\url{https://github.com/GaloisInc/ambient-verifier}.

\paragraph{\pate{} (\textbf{P}atches \textbf{A}ssured up to \textbf{T}race
\textbf{E}quivalence)}

A relational
verification~\citep{relationalVerification1,relationalVerification2} tool that
proves that a property about a binary's observable behavior holds before and
after a patch is applied to the binary's code.
A typical use case for \pate{} is verifying that security-oriented patches fix
undesirable behaviors in a binary without adversely affecting other parts of
the binary.
First, \pate{} splits the original and patched binaries into
\emph{conflict-free acyclic regions} (CFARs).
A CFAR is a collection of basic blocks with control flow between them, but with
no backedges.
\pate{} uses the original binary as the behavioral specification, inferring
frame conditions for each CFAR.
\pate{} then symbolically executes each CFAR in the original and patched
binary, and if the patched binary fails to satisfy the frame condition inferred
from the original binary, then \pate{} produces a summary of the conditions
that lead to exhibiting different behavior.

The source code for \pate{} is publicly available at
\url{https://github.com/GaloisInc/pate}.

\paragraph{\cerridwen}

A tool that quantifies similarities among \xEightySix{} binaries.
The algorithm for ranking similarity largely follows the description in
~\citep{reoptimizedBinarySimilarity}.
\cerridwen{} decomposes a binary into strands---that is, data-flow slices of
basic blocks~\citep{strands}---which are used as units of measurement for
comparing how similar the binary is to a corpus of known binaries.
The basic blocks resulting from \macaw's code discovery are used to compute
each strand, which consists of the operations that lead to the computation of a
single value within a basic block.

The source code for \cerridwen{} is publicly available at
\url{https://github.com/GaloisInc/cerridwen}.

\paragraph{\renovate}

A static binary rewriting library for ELF binaries.
Using \renovate, one can add, remove, or rewrite basic blocks in a statically
linked binary without needing to execute it.
The core of \renovate's rewriting machinery is architecture-independent, and
there exist \xEightySix, PowerPC, and AArch32 backends.
One particular use case for \renovate{} is achieving binary diversity
~\citep{softwareDiversity}---that is, randomizing the layout of basic blocks
to make it more resilient to attackers.

The source code for \renovate{} is publicly available at
\url{https://github.com/GaloisInc/renovate}.

\paragraph{\mcTrace}

An \xEightySix{} and PowerPC binary instrumentation tool that is
heavily inspired by DTrace~\citep{dtrace1,dtrace2}, but does not require any
special kernel support or recompiling the program to be traced.
\mcTrace{} reads the same tracing specifications as DTrace (written in the D
language) and leverage's \renovate's binary rewriting capabilities to insert
probes into binaries.
The traces that these probes generate can support a variety of different
systems.
For example, \mcTrace-instrumented binaries have been successfully run on bare
metal using an MPC5777C PowerPC microcontroller.

The source code for \mcTrace{} is publicly available at
\url{https://github.com/GaloisInc/mctrace}.

\paragraph*{\surveyor}

A library for interactively debugging LLVM, JVM,
\xEightySix, and PowerPC programs, using an interface inspirted by
\lstinline|gdb| and \lstinline|emacs|.
Notably, \surveyor{} is a \emph{symbolic} debugger: it supports setting
breakpoints during \crucible's symbolic execution, and it can inspect the
values of arguments to a basic block, which may possibly be symbolic
themselves.
\surveyor{} also supports stepping through a binary's execution and recording a
trace of symbolic events, which can be replayed later.

The source code for \surveyor{} is publicly available at
\url{https://github.com/GaloisInc/surveyor}.

\section{Related Work}\label{sec:related-work}

\paragraph{Reverse Engineering Frameworks}

The closest tools to \macaw{} in terms of capabilities are industrial binary
analysis frameworks such as Ghidra~\citep{ghidra}, Binary Ninja~\citep{binja},
Radare2~\citep{radare2}, and IDA Pro~\citep{idaPro}.
Like \macaw, these frameworks handle a multitude of binary analysis tasks such
as reverse engineering, disassembly, decompilation, and static analysis.
Each framework also includes a native IR that abstracts some low-level
architectural details---for example, Ghidra's P-Code~\citep{pcode} and
others~\citep{bnil,esil,idaMicrocode}.

The main difference between these tool and \macaw{} lies in their respective
design goals.
Ghidra and similar are intended for industrial-strength reverse engineering
applications.
\macaw{} is intended to be a research-oriented toolkit for rapidly
building novel binary analysis tools.
For the user, these binary analysis frameworks offer a polished
experience which \macaw{} does not seek to provide. For example,
Ghidra features a full, graphical IDE that allows users to visualize
disassembled programs, label fields, rename variables, and more.
We have experimented with different UI approaches via the \surveyor{} tool, but
as exploratory tools, not fully-developed capabilities.

\macaw's greatest strength is providing a platform to build new binary analysis
tools from scratch.
Frameworks such as Ghidra typically support plugins via some API, but this is
usually meant for extending the existing functionality rather than building
standalone applications.
In this sense, \macaw{} occupies a different niche in the design space.
Indeed, it is possible to build a custom analysis tool with \macaw{} and
integrate its capabilities into Ghidra's IDE.

% TODO: have we done this???

Another difference is \macaw's intrinsically typed IR, which encodes invariants
about well-formedness at the type level.
This guarantees that IR values are valid in the context of the analyses they
want to perform.
To our knowledge, none of the other IRs mentioned earlier employ a comparable
approach to IR validity.
There have been efforts to give formal semantics to P-Code, but this effort is
external to the actual P-Code representation used in Ghidra~\citep{pcode}.

\paragraph{Binary Lifting}

\reopt{} is one tool in the large space of binary lifters and rewriters;
see~\citet{schulte} for a comparison.
\reopt{} lifts machine code to the LLVM IR, which puts it in the same category
as lifters such as McSema~\citep{mcSema}, SecondWrite~\citep{secondWrite},
RevGen~\citep{revGen}, and rev.ng~\citep{rev.ng}.
It differs primarily in the style of LLVM code that it produces.
Most LLVM-based lifters map processor registers directly to LLVM variables with
an explicit stack, which results in atypical LLVM code.
\reopt{} unifies the machine code stack with the LLVM stack in an effort to
produce more ``human-friendly'' LLVM.

\paragraph{Symbolic Execution of Machine Code}

\macaw{} leverages \crucible's support for forward symbolic
execution~\citep{symbolicExecution,forwardSymbolicExecution} to analyze the
behavior of programs.
Other binary analysis frameworks use symbolic execution as well, examples of
which include Angr~\citep{angr}, BAP~\citep{bap}, Triton~\citep{triton},
Mayhem~\citep{mayhem}, and KLEE-Native~\citep{kleeNative}.
Most of these leverage \emph{dynamic} symbolic execution (also known as
\emph{concolic} execution), while \macaw{} uses \emph{static} symbolic execution
by lifting a binary to a high-level IR and analyzing it using symbolic values.

This reflects a difference in the types of problems we have typically targeted
with \macaw{}.
Other tools typically use symbolic execution in bug-finding (for which dynamic
symbolic execution is well-suited) while we have often used \macaw{} to support
exhaustive exploration or formal verification (for which static symbolic
execution is necessary).

Angr is static insofar as it lifts binaries to an IR and performs additional
analyses, but it is also provides dynamic analyses such as a symbolic-assisted
fuzzer~\citep{driller}.
This is similar to our \macaw{}-based tool \ambient{}, but Angr uses a
combination of static and dynamic symbolic execution depending on the use case.
For example, Angr can directly interact with files and the operating system,
whereas \ambient{} can only model filesystem and OS interactions in an abstract
way.

Another important distinction is that Angr often concretizes certain types of
symbolic values during execution, depending on the \emph{concretization
strategy} that the user chooses.
With the default concretization strategy, Angr will concretize symbolic
addresses that are written to, as well as symbolic length values passed to the
\lstinline|read()| and \lstinline|write()| system calls.
This is done primarily for performance reasons, as preserving all symbolic
values can often lead to VCs that are prohibitively expensive to solve for.
By contrast, \ambient{} (and \macaw{} more generally) propagates symbolic
values throughout its analyses.
This means that \ambient{} will explore more paths by default, but perhaps at
the expense of additional analysis time.

\paragraph{Formal Verification of Machine Code}

Our \macaw{}-based verification tool SAW is similar to
Vale~\citep{vale1,vale2,vale3}, a tool for writing verified \xEightySix{}
assembly code.
Vale integrates with Low*~\citep{lowStar} to allow verification of mixed C/\xEightySix{} code.
Vale and Low* are used in the implementation of EverCrypt~\citep{everCrypt}, a
library of verified C/\xEightySix{} cryptography.

Unlike Vale and Low*, SAW is primarily used to verify \emph{pre-existing}
C/\xEightySix{} code, whereas Vale and Low* are primarily meant for writing code
and proofs simultaneously.
Because SAW verifies \xEightySix{} code as-is, SAW makes essential use of
\macaw's binary lifting as part of the overall verification process.
Vale, on the other hand, generates \xEightySix{} code from a verified
domain-specific language, so it does not rely on binary lifting.

\section{Conclusions and Next Steps}\label{sec:future-work}

\macaw{} has been a keystone technology for us in developing binary analysis
tools over a decade.
\macaw{}'s core IR builds in multiple features that let us build rapidly and
with confidence, while its symbolic execution capabilities have enabled a
diverse range of testing and verification tools.

Our current focus is on improving \macaw's code discovery capabilities.
A common pain point is discovering the targets of calls through jump
tables---by default, \macaw{} uses simple pattern-based heuristics.
We have developed \macawRefinement{} (\Cref{sec:macaw-refinement}) as a more
accurate alternative. 
However, its reliance on SMT solving can increase discovery time in a way that
is unacceptable for many applications.
We are currently investigating how to limit SMT calls to only the places that
the pattern-based heuristic fails.
Our eventual goal is to make \macawRefinement{} fast enough to be the default
code discovery algorithm which would bring many more target binaries within
reach.

\begin{acks}
This work relates to Department of Navy award \#N00014-22-1-2588 issued by the
Office of Naval Research.
Any opinions, findings, and conclusions or recommendations expressed in this
material are those of the author(s) and do not necessarily reflect the views of
the Office of Naval Research.

\end{acks}

\bibliographystyle{ACM-Reference-Format}
\bibliography{bibliography}

\appendix

% Comment out the lines below to produce a version of the paper without the
% Appendix sections, but maintaining earlier \Cref references in the paper.
% Make sure to also uncomment the \nofiles line at the top of the paper.
%
% On the flip side, to produce a version of the paper with ONLY the Appendix
% sections, comment out the \maketitle and all non-Appendix `\input`s.
% \input{macaw-design-appendix}
% \input{reopt-appendix}

\end{document}